\documentclass{aa}  

\usepackage{array}
\usepackage{times}
\usepackage{mathptmx}
\usepackage[T1]{fontenc}
\usepackage{ae,aecompl}
\usepackage{lscape}
\usepackage{amsmath}	
\usepackage{amssymb}
\usepackage{bm}
\usepackage{graphicx}
\usepackage{xcolor}
\usepackage{txfonts}
\usepackage{hyperref}
\usepackage{bm}
\newcommand{\Msun}{\ensuremath{\,{\rm M}_\odot}}                  
\newcommand{\Rsun}{\ensuremath{\,{\rm R}_\odot}}                  
\newcommand{\Rjup}{\ensuremath{\,{\rm R}_{\rm Jup}}}              
\newcommand{\prism}{\texttt{PRISM}}	
\newcommand{\jktebop}{\texttt{JKTEBOP}} 
\newcommand{\Rearth}{\ensuremath{\,{\rm R}_\oplus}}		    

\begin{document}

   \title{Simulations of starspot anomalies within TESS exoplanetary transit light curves -- I.}
\titlerunning{Simulations of starspot anomalies within TESS}
   \subtitle{The detection limits of starspot anomalies in TESS light curves}

   \author{J. Tregloan-Reed
          \inst{1}
           \and
          E. Unda-Sanzana\inst{1}
          }
\authorrunning{Tregloan-Reed 
\and 
Unda-Sanzana
}

   \institute{Centro de Astronom\'{i}a (CITEVA), Universidad de Antofagasta,
              Avenida U. de Antofagasta 02800, Antofagasta, Chile\\
              \email{jeremy.tregloanreed@uantof.cl}
             }

   \date{Received April 19,2019; accepted August 14,2019}

 
  \abstract
   {The primary targets of the NASA {\it Transiting Exoplanet Survey Satellite} will be K and M dwarf stars within our solar neighbourhood. Young K and M dwarf stars are known to exhibit a high starspot coverage ($\approx50$\,\%), however, older stars are known to show fewer starspots. This implies that TESS 2\,min cadence transit light curves may contain starspot anomalies, and if so, will require transit-starspot models to accurately determine the properties of the system.}
  {The goals are to determine if starspot anomalies can manifest in TESS transit light curves, to determine the detection limits of the starspot anomalies and to examine the relationship between the change in flux caused by the starspot anomaly and the planetary transit.}
   {20573 simulations of planetary transits around spotted stars were conducted using the transit-starspot model, \prism. In total 3888 different scenarios were considered using three different host star spectral types, M4V, M1V and K5V. The mean amplitude of the starspot anomaly was measured and compared to the photometric precision of the light curve, to determine if the starspot anomaly's characteristic ``blip'' was noticeable in the light curve. }
   {The simulations show that, starspot anomalies will be observable in TESS 2\,min cadence data. The smallest starspot detectable in TESS transit light curves has a radius of $\approx1900$\,km. The starspot detection limits for the three host stars are: $4900\pm1700$\,km (M4V), $13800\pm6000$\,km (M1V) and $15900\pm6800$\,km (K5V). The smallest change in flux of the starspot ($\Delta F_\mathrm{spot} = 0.00015\pm0.00001$) can be detected when the ratio between the planetary and stellar radii, $k = 0.082\pm0.004$. }
  {The results confirm known dependencies between the amplitude of the starspot anomaly and the photometric parameters of the light curve. The results allowed the characterisation of the relationship between the change in flux of the starspot anomaly and the change in flux of the planetary transit for TESS transit light curves.}


   \maketitle
%

\section{Introduction} \label{Sec:1}

The next generation of transit planet hunters (e.g.,\ Next Generation Transit Survey: \citealt{NGTS}; NASA {\it Transiting Exoplanet Survey Satellite}: \citealt{TESS,TESS2,TESS3}) will observe transiting planets orbiting both K and M dwarf stars, within, the solar neighbourhood (with an emphasis on super-Earth:\,1.25\Rearth to 2\Rearth; sub-Neptune:\,2\Rearth to 4\Rearth; Neptune:\,4\Rearth to 6\Rearth class planets). TiO absorption bands show that young K dwarfs can have 20-40\% starspot coverage \citep{Oneal2004}, while young M dwarfs can have a starspot coverage of 40\,$\pm$\,10\,\% \citep{Jackson2013}. Older stars are known to show fewer starspots, however, the dependence on age has not been reliably quantified. This leads to the potential of an increased likelihood
that the transits observed by these new planet hunters, will contain starspot anomalies. 

Late-type stars (G, K, M) have magnetic fields, which often manifest themselves as dark spots on the stellar surface. When a transiting planet passes in front of a starspot, a short-term increase in the brightness of the star is seen because the planet is temporarily blocking a fainter part of the star's surface \citep{Silva2003}. This ``blip'' in the light curve can be modelled to determine the size and position of the starspot (e.g., WASP-19: \citealt{Jeremy2012, Mancini2013b}; WASP-6: \citealt{Jeremy2015}; CoRoT-2: \citealt{Silva2010b}; WASP-4: \citealt{Sanchis2011a}; HAT-P-11: \citealt{Sanchis2011b}; Kepler-30: \citealt{Sanchis2012}; Kepler-63: \citealt{Sanchis2013}; HATS-2: \citealt{Mohler2013}; Qatar-2: \citealt{Mancini2014}; WASP-41: \citealt{Sou2016}; WASP-52: \citealt{Mancini2017}; GJ\,3470: \citealt{Chen2017}: Kepler-17: \citealt{Valio2017}). 

Starspots can affect the shape of the light curve \citep{Silva2010} and if not correctly modelled can lead to biased measurements of the system parameters (e.g., planetary radius: \citealt{Nikolov2013}; limb darkening coefficients: \citealt{Ballerini2012}; time of minimum light: \citealt{Sanchis2011a}). However, if the perturbations are accounted for then an accurate measurement of the planetary radius will be achievable, improving investigations into the atmosphere, structure and evolution of a planet \citep{Fortney2007}. 

When two closely spaced transit light curves, both of which contain a starspot anomaly (due to the same starspot), are accurately modelled, the position of the starspot can be determined at two distinct points in time, allowing an extraction of the latitudinal stellar rotation period ($P_*$) and sky-projected orbital obliquity (the sky projection of the angle between the orbit of the planet and the spin of the host star), $\lambda$ from the transit light curves \citep[e.g.,][]{Sanchis2011a,Sanchis2011b,Jeremy2012,Jeremy2015,Sou2016,Mancini2017}.

Over the years there have been multiple transit-starspot models developed both by the eclipsing binary star community (e.g., {\tt Wilson-Devinney} code: \citealt{wdcode,Wilson1979,Wilson1990,Wilson2008,Wilson2012}; {\tt PHEOBE}: \citealt{phoebecode,phoebecode2}) and the exoplanet community (e.g., \prism: \citealt{Jeremy2012,Jeremy2015,Jeremy2018}; {\tt SOAP-T}: \citealt{Oshagh2013}; {\tt spotrod}: \citealt{Beky2014}; {\tt KSint}: \citealt{Montalto2014}; {\tt ellc}: \citealt{Maxted2016}; {\tt StarSim}: \citealt{Herrero2016}; {\tt PyTranSpot}: \citealt{Juven2018}). With a large variety of models readily available to the community, the treatment of starspot anomalies and their associated effects on light curves, no longer cause difficulties or delays in the analysis of the exoplanetary light curves.

Locating starspot anomalies in light curves of smaller planets requires a significantly higher precision in the light curve because the size of the starspot anomaly scales linearly with the area of the planetary disc \citep{Jeremy2012}. This precludes the use of ground-based data for small planets, but the remarkable quality of the light curves from dedicated space missions such as the {\it Kepler} satellite \citep{Kepler} makes this work viable. Whilst the temporal sampling of {\it Kepler} data (30\,min cadence) is too low to resolve starspot anomalies for the vast majority of the stars observed, the NASA {\it Transiting Exoplanet Survey Satellite} (TESS) will observe $\ge$\,200,000 stars at a 2\,min cadence \citep{Stassun2017} and provide a large set of light curves suitable for detecting and measuring starspots.

TESS will spend two years looking for super-Earth planets ($<$\,2\Rearth) transiting between 200,000 and 400,000 pre-selected stars at a cadence of two minutes, \citep{Stassun2017} with a particular emphasis on M dwarf stars. Recent simulations for TESS indicate the potential to detect 1700 planets from the smallest selection of pre-selected stars \citep{Sullivan2015, Sullivan2015_erratum}. With 1100 of these planets predicted to be sub-Neptune (2\Rearth to 4\Rearth) class planets, while, 419 super-Earth or smaller ($<$\,2\Rearth) planets, are expected to be found orbiting M dwarfs. With a further 137 super-Earth or smaller planets orbiting F, G, K stars \citep{Sullivan2015, Sullivan2015_erratum}. TESS is already exceeding early expectations with 21 confirmed planets (e.g., Pi Mensae\,c: \citealt{Huang2018}: LHS 3884\,b; \citealt{Vanderspek2018}: HD 21749\,b; \citealt{Dragomir2019}: HD 202772A\,b; \citealt{Wang2019}) with a further 336 TESS Objects of Interest (ToIs)\footnote{From \href{https://archive.stsci.edu/prepds/tess-data-alerts}{TESS data MAST page}; accessed 2019-06-24.}.

The objective of this work is to determine the detection limits of starspot anomalies in TESS exoplanetary transit light curves. This is accomplished by varying the variables that directly or indirectly alter the duration and\,/\,or amplitude of the starspot anomaly in the light curve, and compare the results against the photometric noise of the light curve: starspot size ($r_\mathrm{spot}$) and temperature ($T_\mathrm{spot}$), host star radius ($R_\mathrm{*}$) and effective temperature ($T_\mathrm{eff}$), planetary radius ($R_\mathrm{p}$), semi-major axis ($a$), orbital period ($P$), orbital inclination ($i$), observing cadence and the frequency ($\nu$) of the observation. Through this, it will be possible to determine the physical properties of starspots at the detection limits for TESS transit light curves.

The outline of this paper is as follows. Section\,\ref{Sec:2} describes how the contrast of the starspots was calculated in the TESS passband, for use in the simulations and describes which parameters influence the size and shape of the starspot anomalies. The section then describes how the simulations were designed using the \prism\ model, to produce the simulated light curves. Section\,\ref{Sec:3} presents the results, separated into three host star spectral types; K5V dwarf, M1V and M4V dwarf stars. Section\,\ref{Sec:4} discusses the results and gives the overall conclusions to the simulations.

\section{Simulating starspot anomalies with \prism} \label{Sec:2}

The transit-starspot model \prism\ \citep{Jeremy2012,Jeremy2015,Jeremy2018} was used to perform the simulations in this work. \prism\footnote{The latest version of \prism\ is available from \href{https://github.com/JTregloanReed/PRISM_GEMC}{GitHub}}. is written in \textsc{idl}\footnote{For further details see \href{http://www.harrisgeospatial.com/ProductsandTechnology/Software/IDL.aspx}{\tt http://www.harrisgeospatial.com/Prod\\uctsandTechnology/Software/IDL.aspx}\,.} (Interactive Data Language) and uses a pixellation approach to model the stellar disc in a two-dimensional array, by subdividing the star into many individual elements. These elements can then be described by a two-dimensional vector in Cartesian coordinates. Each element is then assigned an intensity value based on if a stellar feature is present at that location and then applying the quadratic limb darkening law over the entire stellar disc. The planet is then set to transit the star. For each data point in the transit light curve, the total received intensity is calculated based on which elements of the star are visible. The pixellation method is an ideal method for introducing stellar features to the stellar disc because the model allows individual intensities to be assigned to individual elements at specific coordinates \citep[see][]{Jeremy2012}. This allows researchers to use \prism\ to model both occulted and unocculted starspots, and so, help ``rule-out'' starspot trends when examining wavelength dependent transit depth (planetary radii) variations. 

The original version of \prism\ used six parameters to model the light curve; ratio between the planetary and stellar radii\,($k$), sum of the fractional ($r_\mathrm{p} = R_\mathrm{p} / a$ and $r_\mathrm{*} = R_\mathrm{*} / a$) planetary and stellar radii\,($r_\mathrm{p} + r_\mathrm{*}$), linear and quadratic coefficients of the quadratic limb darkening law\,($u_1$ and $u_2$), orbital inclination\,($i$) and time of minimum light\,($T_0$). Combined with four additional parameters for each starspot; longitude of the spot's centre\,($\theta$), co-latitude of the spot's centre\,($\phi$), angular size\,($r_\mathrm{spot}$) and contrast, the surface brightness of the starspot versus the immaculate photosphere\,($\rho_\mathrm{spot})$ \citep{Jeremy2012}. Since its publication, \prism\ has had two major updates. The first added two orbital parameters; orbital eccentricity\,($e$) and argument of periastron\,($\omega$) \citep{Jeremy2015}. The second major update included a third light parameter and the out-of-transit detrending polynomial coefficients \citep{Jeremy2018}.

During its development \prism's ability to model the stellar disc and transit was tested by using \jktebop\ \citep{Southworth2004, Southworth2005, Southworth2007b, Southworth2008, Southworth2009, Southworth2010, Southworth2011, Southworth2013} as a benchmark transit modelling program. For this a series of light curves were generated by both \prism\ and \jktebop\ using the same transit parameter values. The average difference between the two models for all the tests was $\sim10$\,ppm, six times lower than the expected noise floor in TESS data \citep[see][]{Sullivan2015}.

Since the development of \prism\, other researchers have used the model to ascertain the photometric parameters of a transiting system and derive the parameters of the detected starspots observed in transit light curves \citep[e.g.,][]{Mancini2013b, Mancini2014, Mohler2013, Chen2017}. \prism\ has also been used to help calibrate and benchmark other transit-starspot models (e.g., {\tt spotrod}: \citealt{Beky2014}; {\tt PyTranSpot}: \citealt{Juven2018}).

\prism\ measures $r_\mathrm{spot}$ in units of angular radius ($^\circ$). Consequently, an $n^\circ$ starspot maintains a constant ratio between the stellar and starspot surfaces, for all values of $R_\mathrm{*}$. The simulations in this work use three different host stars, and as such, three different $R_\mathrm{*}$. To avoid the projection effects on the spherical surface of the star we quote the starspot stellar surface radius, in units of $R_\mathrm{*}$, where:

\begin{equation}
 n^\circ \equiv n^\circ\left(\frac{\pi}{180^\circ}\right)R_\mathrm{*} \ ,
\end{equation}

\noindent where a starspot with an angular radius of $1^\circ$ has a stellar surface radius of $0.017R_\mathrm{*}$ 

\subsection{Typical starspot contrast in the TESS passband} \label{Sec:2.1}

While the temperature of the starspot ($T_\mathrm{spot}$) is fixed across all wavelengths, $\rho_\mathrm{spot}$ is not, it is dependent on the frequency of the observation, $\nu$. At bluer wavelengths $\rho_\mathrm{spot}$ reduces (higher contrast), increasing the amplitude of the starspot anomaly. This is because a starspot is cooler than the surrounding photosphere and both the starspot and photosphere can be treated as black bodies \citep{Rabus2009, Sanchis2011b}. When the observation wavelength shifts towards a redder wavelength, $\rho_\mathrm{spot}$ increases (lower contrast), and so, the amplitude of the starspot anomaly can potentially fade into the observational noise.

Assuming that both the starspot and surrounding photosphere radiate as blackbodies, \citet{Silva2003} uses the ratio between the intensities of the starspot and surrounding photosphere to give an equation to find $\rho_\mathrm{spot}$ from the stellar effective temperature ($T_\mathrm{eff}$), $T_\mathrm{spot}$ and $\nu$. By doing so, \citet{Silva2003} incorporates the spectral signature of the starspot into $\rho_\mathrm{spot}$:

\begin{equation} \label{eq.1}
 \rho_\mathrm{spot} = \frac{\exp\left(h\nu / k_\mathrm{B} T_\mathrm{eff}\right) - 1}{\exp\left(h\nu / k_\mathrm{B} T_\mathrm{spot}\right) - 1} \ ,
\end{equation}

\noindent where $h$ is Planck's constant and $k_B$ is Boltzmann constant.

TESS will use its own $T$ (TESS) passband with a spectral response between 600--1000\,nm with the Johnson-Cousins $I_C$ passband lying near the centre of the $T$-passband \citep{Sullivan2015}. The effective wavelength midpoint of the $I_C$ passband lies approximately at 785\,nm. Because a large proportion of the emitted flux from an M dwarf lies within the optical red--infrared (potentially engulfing the starspot signal due to the redder observational wavelengths) Eq.\,\ref{eq.1} was used to calculate the contrast for a typical starspot on a cool M dwarf at 600\,nm, 785\,nm and 1000\,nm. Using a $T_\mathrm{eff} = 3200$\,K (transit detection limit in $T$ passband; \citealt{Sullivan2015}) and setting $T_\mathrm{spot}$ to be 200\,K cooler than the photosphere (i.e., $T_\mathrm{spot} = 3000$\,K), in-line with starspot temperatures of fully convective M dwarf stars \citep[e.g.,][]{Barnes_J2015}, $\rho_\mathrm{spot}$ was calculated to be $\rho_\mathrm{spot} = 0.60$, $\rho_\mathrm{spot} = 0.68$ and $\rho_\mathrm{spot} = 0.74$ for the three respective wavelengths. The amplitude of a starspot anomaly is primarily dependent on $r_\mathrm{spot}$ and $\rho_\mathrm{spot}$. However, there will be a critical threshold for $\rho_\mathrm{spot}$ where the starspot anomaly will no longer be detected, irrespective of $r_\mathrm{spot}$. Previous work using ground-based telescopes, have found starspot anomalies with $\rho_\mathrm{spot} > 0.74$ (e.g., HAT-P-36, $\rho_\mathrm{spot} = 0.92$: \citealt{Mancini2015}; WASP-41, $\rho_\mathrm{spot} = 0.90$: \citealt{Sou2016}; WASP-6, $\rho_\mathrm{spot} = 0.80$: \citealt{Jeremy2015}; WASP-19, $\rho_\mathrm{spot} = 0.78$: \citealt{Jeremy2012}). This indicates that $\rho_\mathrm{spot}$ of a typical starspot anomaly lying within the photosphere of a M dwarf star, observed in the $T$-passband, is below the critical threshold to be detected by ground-based telescopes and implies that starspot anomalies should be detected in TESS transit light curves.

\subsection{Parameter impact on the shape of a starspot anomaly} \label{Sec:2.2}

Before we begin simulating starspot anomalies in transit light curves, we need to examine the parameters which directly or indirectly influence the size (amplitude) and shape (duration) of the anomaly. The parameters can be divided into four groups.

\begin{itemize}
\item Starspot parameters.
\item Stellar and planetary parameters.
\item Orbital parameters.
\item Observing constraints
\end{itemize}

The first group of parameters belong to the starspot itself $\theta$, $\phi$, $r_\mathrm{spot}$ and $\rho_\mathrm{spot}$ (or $T_\mathrm{spot}$). The amplitude of a starspot anomaly is directly proportional to $\rho_\mathrm{spot}$ and the surface area of the starspot \citep{Jeremy2012}. As $\rho_\mathrm{spot}$ decreases (i.e., the starspot darkens / cools), $r_\mathrm{spot}$ must reduce, to maintain the amplitude of the starspot anomaly. In essence, a small starspot, will need a large ratio between $T_\mathrm{spot}$ and $T_\mathrm{eff}$, to be detected (see Eq.\,\ref{eq.1}). The location of the starspot, $\theta$ and $\phi$ have an indirect effect on the amplitude and duration of the anomaly. When the position of the starspot approaches the limb of the star, there is a reduction in the 2\,D projection of the spot's surface area, due to foreshortening along the radial vector (the vector between the centres of the stellar disc and starspot). Hence, reducing the amplitude of the starspot anomaly. As the radial vector, rotates (by altering the Azimuth angle) to being parallel with the transit cord, the anomaly's duration will be reduce too.

The next group of parameters is related to the host star; $T_\mathrm{eff}$, $R_\mathrm{*}$ and the planet; $R_\mathrm{p}$. It was shown in both Section\,\ref{Sec:2.1} and Eq.\,\ref{eq.1} how $T_\mathrm{eff}$ directly alters $\rho_\mathrm{spot}$ thereby, indirectly influencing the amplitude of the anomaly. The amplitude of a starspot anomaly is also directly proportional to $k$, the ratio between $R_\mathrm{p}$ and $R_\mathrm{*}$ \citep{Jeremy2012}. If $k$ is too small in relation to the noise then the change in flux due to the planet occulting the starspot, will not be detectable over the noise in the light curve.

The orbital parameters $i$, $a$ and $P$ all indirectly impact the amplitude and duration of the starspot anomaly. $i$ controls the impact parameter\,($b$) and therefore, $i$ controls the stellar latitude at which the transit cord will cross. Because occulted starspots must lie on the transit chord to be observed then as $i$ decreases, the position of the starspot will move towards the stellar pole (i.e., the limb of the stellar disc). By doing so, the 2\,D projection of the starspot's surface area will decrease due to foreshortening. $b$ is also controlled by $a$ combined with the orbital eccentricity\,($e$) and the argument of periastron\,($\omega$) \citep[e.g.,][]{Winn2010Book}:  

\begin{equation} \label{eq.2}
b = \frac{a \cos i}{R_\mathrm{*}} \left(\frac{1-e^2}{1+e\sin\omega}\right) \ .
\end{equation}

\noindent Because $P$ is related to $a$ via Kepler's third law and directly controls the transit duration \citep[e.g.,][]{Winn2010Book}, it becomes apparent how $a$ and $P$ can both directly control the duration and indirectly influence the amplitude of the starspot anomaly.

The final group consists of just two parameters $\nu$ and the observing cadence. In Section\,\ref{Sec:2.1} it was shown how $\nu$ alters the amplitude of the starspot anomaly. Where the amplitude of the starspot anomaly increases towards bluer wavelengths and reduces towards redder wavelengths. The observing cadence though, does not physically affect the size or shape of the starspot anomaly. However, it does alter our perception of the size and shape. When the cadence decreases (from high to low cadence), the number of data points that describe the starspot anomaly will decrease, altering our perception of the anomaly’s true shape.  This will increase the uncertainty in the anomaly's duration, and so, increase the uncertainty in the measured parameters which influences the duration of the starspot anomaly.

Because these simulations are based on TESS transit light curves, the observational cadence is fixed at 2\,min \citep{Stassun2017} and $\nu$ is fixed at $600$\,nm, $785$\,nm \& $1000$\,nm \citep{Sullivan2015}. Consequently both the cadence and $\nu$ are considered as constraints and are used in the simulations, at fixed values.

\subsection{Simulating starspots in TESS transit light curves} \label{Sec:2.3}

The primary goal of this work is to find the detection limits of starspot anomalies appearing within TESS exoplanet light curves. TESS was developed to find small planets ($<$\,6\Rearth) orbiting local K and M dwarf stars \citep{TESS}. Due to this, the host star and planetary properties used in these simulations were based on the selections used by \citet{Sullivan2015,Sullivan2015_erratum}, covering K \& M late-type stars with small planets, $<$\,6\Rearth.

When predicting the planetary yields of TESS \citet{Sullivan2015} calculated that TESS will have a photometric sensitivity of 200\,ppm for an $I_C$ 10$^{th}$ magnitude star. Because of this, the light curves generated by \prism\ had added Gaussian noise: $60\pm3$\,ppm\footnote{Mission specified noise floor at 60\,ppm\,hr$^{1/2}$ \citep[see][]{Sullivan2015}}, $100\pm5$\,ppm, $150\pm7.5$\,ppm and $200\pm10$\,ppm. By doing so, allows us to simulate the photometric sensitivity of light curves for host stars brighter than $I_C = 10$ (after undergoing the TESS data processing pipeline and known systematics removal, see \citealt{SPOC2016}). 

With the TESS observations being performed in the $T$-passband (600\,nm--1000\,nm; \citealt{Sullivan2015}) the simulations were run at three wavelengths: 600\,nm, 785\,nm and 1000\,nm. These wavelengths were selected to give a good average approximation to the $T$-passband (785\,nm) and to obtain results at the bluest (600\,nm) and reddest (1000\,nm) edge of the $T$-passband spectral response, and so, produce the largest and smallest amplitudes for the starspot anomalies. 

Examining the range of parameters described in Section\,\ref{Sec:2.2} it becomes apparent that the parameter space which controls the amplitude and duration of a starspot anomaly, is large and complex due to multiple degeneracies between the parameters (e.g., $r_\mathrm{spot}\leftrightarrow\rho_\mathrm{spot}$, $\phi\leftrightarrow i$, $P\leftrightarrow a$ and $\phi\leftrightarrow P$). However, for this work we are only interested in the detection limits of starspot anomalies found within TESS light curves (i.e., smallest $r_\mathrm{spot}$ for a given scenario). Therefore we can streamline the parameter space by using a mixture of predetermined key parameter values / scenarios (e.g., setting $R_\mathrm{*}$ and $T_\mathrm{eff}$ for typical K \& M dwarf stars) and search for the smallest detectable $r_\mathrm{spot}$.

In the simulations the position of the starspot was set at the centre of the stellar disc, therefore, minimising the effect from foreshortening. As discussed in Section\,\ref{Sec:2.2} foreshortening reduces the amplitude of the starspot anomaly. Therefore, due to the geometric nature of spherical foreshortening, the smallest detectable $r_\mathrm{spot}$ will be for starspots lying at the centre of the stellar disc. This is because when the starspot moves closer to the limb, $r_\mathrm{spot}$ needs to increase, so to maintain the same two dimensional projected surface area of the starspot. Because we want to detect the smallest starspot for a given set of circumstances. Placing the starspot at the centre of the stellar disc, will achieve this. 

\begin{figure} \includegraphics[width=0.48\textwidth,angle=0]{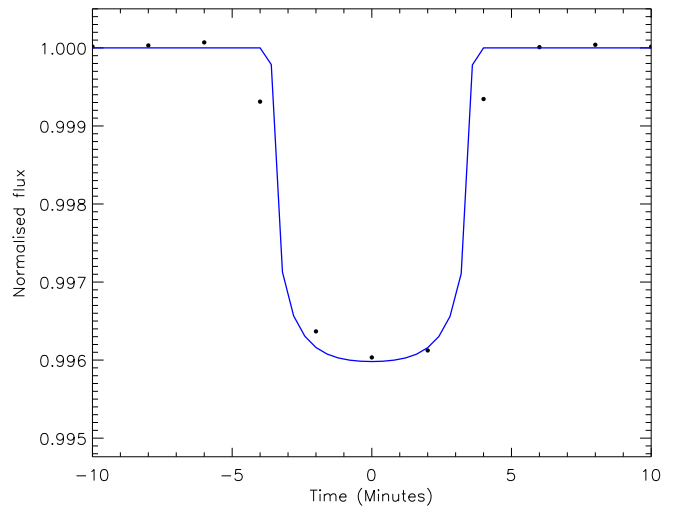} 
\caption{\label{fig:1} Simulated two minute cadence TESS light curve, of a 1\Rearth\ planet orbiting an M4V dwarf star with $P = 12$\,d and an rms scatter of 104\,ppm (generated by \prism). The transit duration is approximately eight minutes and only three data points lie within the transit. } \end{figure}

To determine TESS planetary yields \citet{Sullivan2015,Sullivan2015_erratum} used an M4V dwarf with $T_\mathrm{eff} = $3200\,K, a radius of $R_\mathrm{*} = 0.155$\,\Rsun, solar metallicity and at 1\,Gyr (in keeping with the Dartmouth Stellar Evolution Database \citealt{Dotter2008}). The $T_\mathrm{eff}$ was selected because M dwarfs cooler than 3200\,K, will be too faint for planetary transit detection in the $T$ passband \ \citep{Sullivan2015}. For this work it was therefore natural to start with an M4V dwarf star. The smaller value for $R_\mathrm{*}$ and by association, larger $k$, meant that a 1\,\Rearth\ radius could be used in the M4V dwarf simulations. The planetary radii used, ranged from 1-3\Rearth\ in 0.25\Rearth\ increments. An upper limit of 3\Rearth\ was used due to the rare occurrence of close in ($<$\,50\,d), large ($>$\,3\Rearth) planets orbiting M dwarf stars, determined from population studies using Kepler and K2 data \citep{Dressing2013, Dressing2015, Morton2014}. Though the recent discovery of NGTS-1\,b: an $\approx$\,15\Rearth\ hot-Jupiter transiting an M\,0.5 dwarf star with a 2.65\,d orbital period \citep{Bayliss2018}, along with two previously known close in large planets transiting M dwarfs (Kepler-45\,b; \citealt{Johnson2012}: HATS-6\,b; \citealt{Hartman2015}) appears to be lifting this limit.

A key objective of TESS is to find suitable planets for further study within the habitable zone of the host star. For a planet orbiting an M4V dwarf star with a relative insolation ($S / S_\oplus$) set at the inner ($S / S_\oplus = 1.0$), mean ($S / S_\oplus = 0.6$) and outer edge ($S / S_\oplus = 0.2$) of the habitable zone \citep{Kopparapu2013} will have an orbital period between 12 and 40\,days \citep[Eq.\,13][]{Sullivan2015}. The difficulty in detecting starspot anomalies at these orbital periods is that the transit duration, is less than 10 minutes (see Fig.\,\ref{fig:1}). This results in the duration of a starspot anomaly being less than a minute, which is not favourable when using two minute cadence data. Unlike transit detections, which can use phase--folded light curves to increase the number of data points within the transit, and so, aid in the transit detection, a starspot anomaly can generally only be detected using a single transit, due to the rotation of the host star smearing the different positions of the starspot anomaly when viewing multiple phase--folded light curves. Due to this, the orbital period in the M4V dwarf simulations was set at 1, 2 and 3\,days, to allow sufficient sampling in the single light curves to detect a starspot anomaly. It should be stated though, that a subset of TESS data will be made available with a cadence of less than two minutes (20\,s) for the asteroseismology community \citep{TESS3}.

\begin{table} \centering
\caption{\label{Tab.1} Calculated values of $a$ for each value of $P$, for the three simulated host stars.}
\setlength{\tabcolsep}{8pt} \vspace{-5pt}
\begin{tabular}{cccccc} 
\hline\hline
 \multicolumn{2}{c}{M4V}  &  \multicolumn{2}{c}{M1V}  &   \multicolumn{2}{c}{K5V}     \\    
   $P$\,(d) & $a$\,(AU) &  $P$\,(d) & $a$\,(AU) &  $P$\,(d) & $a$\,(AU)   \\   
\hline
  1.0 &   0.0091 &  2.0 & 0.0242 &  2.0 & 0.0269  \\ 
  2.0 &  0.0144 &  3.0 & 0.0317 &  4.0 & 0.0427  \\ 
  3.0 &  0.0189 &  4.0 & 0.0383 &  6.0 & 0.0560  \\ 
\hline \end{tabular} 
\end{table}

Kepler's third law was used to calculate $a$ from $P$ using a stellar mass, $M_* = 0.1$\,\Msun\ and are given in Table\,\ref{Tab.1}. The stellar mass was determined using $R_\mathrm{*} = 0.155$\,\Rsun\ combined with Eq.\,10 (second order polynomial) from \citet{Boyajian2012}, derived from  interferometric measurements of M\,\&\,K dwarfs. We then calculated $\log g = 5.06$ for the M4V dwarf and combined with $T_\mathrm{eff} = 3200$\,K, determined the quadratic limb darkening coefficients, $u_1$\,\&\,$u_2$ (Table\,\ref{Tab.2}) using grids developed by \citet{Claret2017} using stellar atmosphere models and are designed for use in the TESS passband.

\citet{Barnes_J2015} looked at starspot distributions of two fully convective M dwarfs, M4.5V: GJ\,791.2\,A and M9V: LP\,944-20 and found starspot temperatures around 200\,K to 300\,K cooler than the surrounding photosphere. In-line with this, we select four equidistant starspot temperatures 3000\,K, 3050\,K, 3100\,K and 3150\,K for the 3200\,K M4V dwarf simulations.

\begin{figure} \includegraphics[width=0.24\textwidth,angle=0]{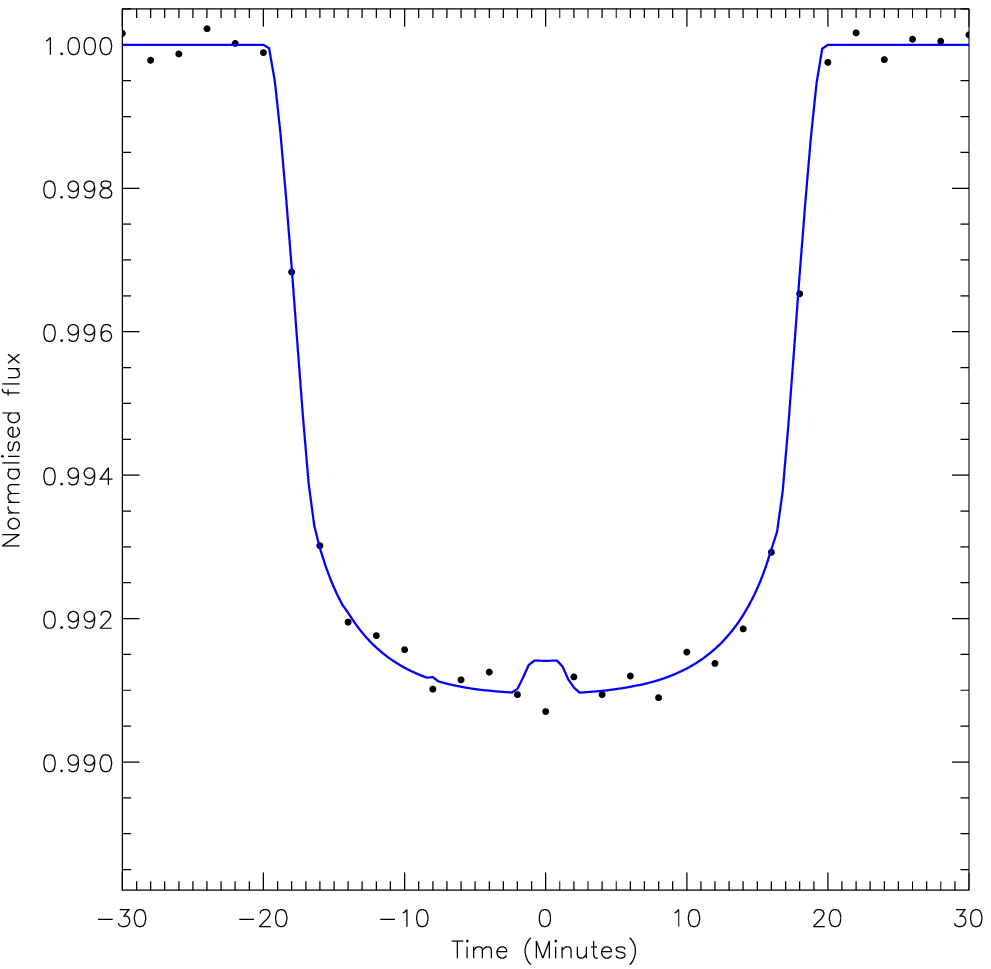} \includegraphics[width=0.23\textwidth,angle=0]{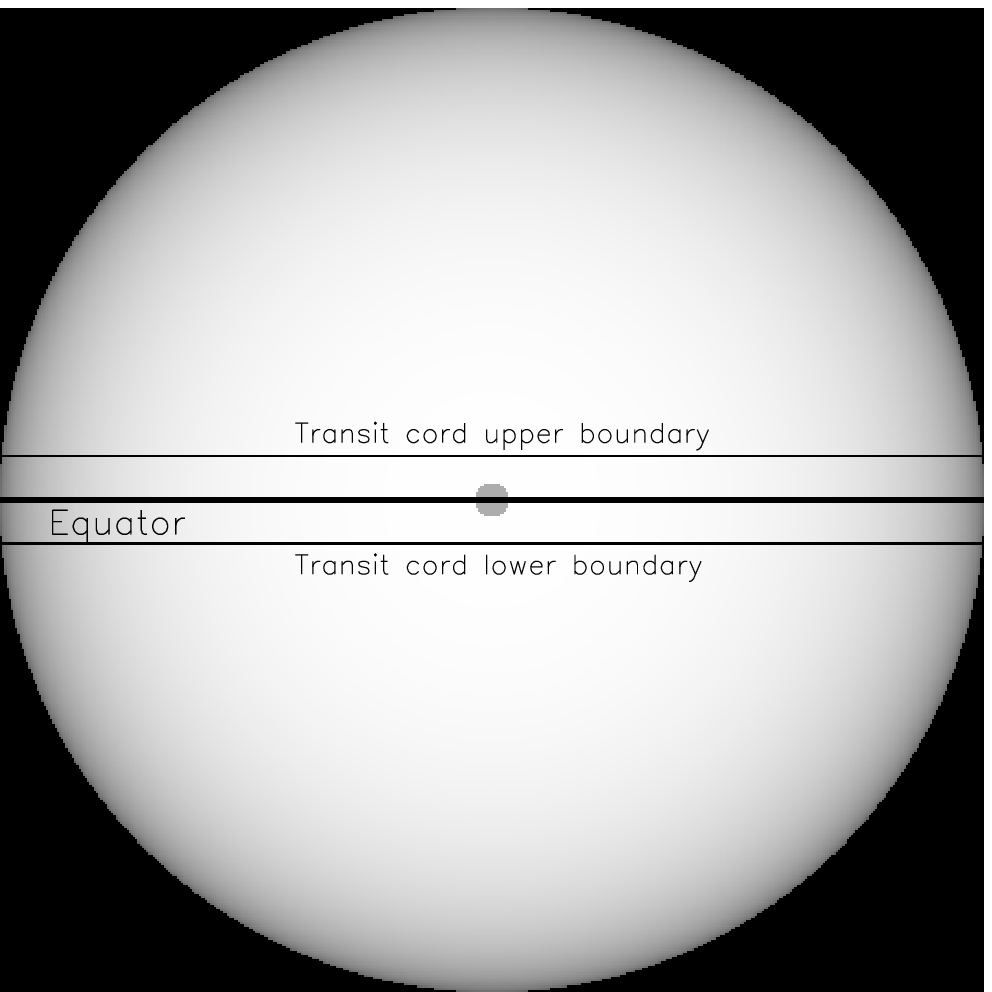} 
\caption{\label{fig:2} Example light curve (left) and stellar disc (right) generated by \prism. Light curve: The solid line represents the noise-free synthetic light curve containing the starspot anomaly, while, the filled circles represent the ``spot-free'' synthetic light curve (see Section\,\ref{Sec:3}) with added Gaussian noise. Stellar disc: The central solid line represents the stellar equator (assuming orbital alignment) and the upper and lower solid lines represent the upper and lower boundaries of the transit cord. These were generated with a 1.5\Rearth\ super-Earth planet transiting a 3200\,K, 0.155\Rsun\ M4V dwarf star with $i =90.0^\circ$ and $P = 1$\,d. The starspot properties are: $\theta = 0^\circ$, $\phi = 90^\circ$, $r_\mathrm{spot} = 0.035R_\mathrm{*}$ and $T_\mathrm{spot} = 3000$\,K. The observational wavelength and the rms scatter for the simulated data were set at 785\,nm and $150\pm7.5$\,ppm respectively. } \end{figure}

1296 different scenarios were used for the M4V dwarf star simulations. Each scenario was derived from alternating the nine $R_\mathrm{p}$, three orbital periods, three observational wavelengths (600\,nm, 785\,nm and 1000\,nm), four starspot temperatures and four noise levels (60\,ppm, 100\,ppm, 150\,ppm and 200\,ppm). $r_\mathrm{spot}$ was then varied to find the detection limit for each of the 1296 scenarios. Fig.\,\ref{fig:2} shows an example light curve and stellar disc from \prism\ for one of the simulations.

Due to the small radius ($R_\mathrm{*} = 0.155$\,\Rsun) of the initial host star, the second host star used in the simulations was a larger M1V dwarf star, with $T_\mathrm{eff} = 3700$\,K \citep[Table\,1][]{Sullivan2015}. This allowed the simulations to examine the detection limit of starspot anomalies at the two ends of the range of $R_\mathrm{*}$ and $T_\mathrm{eff}$ of M dwarfs, observed by TESS. The $R_\mathrm{*}$ and $M_\mathrm{*}$ of the M1V dwarf were extrapolated using the polynomials derived from interferometric measurements of M\,\&\,K dwarfs \citep[Eq.\,8\,\&\,10][]{Boyajian2012} using $T_\mathrm{eff}=3700$\,K. The results\footnote{We ignore the coefficient uncertainties in the polynomials presented by \citet{Boyajian2012} as we are generating hypothetical host stars, whose physical properties are consistent with the physical properties of observed stars of the same spectral type.} are given in Table\,\ref{Tab.2} along with the extrapolated properties of the M4V and K5V dwarf host stars. We employ the same procedure that was used for the M4V dwarf in determining $u_1$ and $u_2$ from \citet{Claret2017} with the results reported in Table\,\ref{Tab.2}.

\begin{table} \centering
\caption{\label{Tab.2} Extrapolated values of $R_\mathrm{*}$ and $M_*$ using $T_\mathrm{eff}$ from polynomials derived from interferometric measurements of M \& K dwarf stars \citep[Eq.\,8\,\&\,10][]{Boyajian2012}, for the three simulated host stars combined with the calculated $\log g$ and $u_1$\,\&\,$u_2$ for the quadratic limb darkening law from \citet{Claret2017}.}
\setlength{\tabcolsep}{6pt} \vspace{-5pt}
\begin{tabular}{lcccccc} 
\hline\hline
 Spectral  & $T_\mathrm{eff}$ & $R_\mathrm{*}$ & $M_*$ & $\log g$ & $u_1$ & $u_2$  \\   
Type  & ($K$) & (\Rsun) & (\Msun) & (cgs) \\
\hline
M4V & 3200  & 0.155 & 0.10 & 5.06 & 0.1533 & 0.4776\\ 
M1V & 3700 & 0.493 & 0.52 & 4.77& 0.1737 & 0.4118\\ 
K5V  & 4100 & 0.623 & 0.65 & 4.66 & 0.3955 & 0.2618\\ 
\hline \end{tabular} 
\end{table}

Due to a threefold increase in the stellar radius, the planetary radii used in the M1V simulations ranged from 2-4\Rearth\ in 0.25\Rearth\ increments. Three orbital periods were selected at 2,\,3\,\&\,4\,days and the corresponding values for $a$ were calculated using Kepler's third law using a stellar mass of $0.52$\,\Msun\ (see Table\,\ref{Tab.2}) and are given in Table\,\ref{Tab.1}. Using an $T_\mathrm{eff} = 3700$\,K for the M1V dwarf star, four equidistant starspot temperatures were selected: 3400\,K, 3475\,K, 3550\,K and 3625\,K, giving a further 1296 different scenarios (Using the same three observational wavelengths and four noise levels that were used in the M4V dwarf simulations) using an M1V dwarf host star. The same procedure that was used in the M4V simulations was used for the M1V simulations. In that for each of the 1296 scenarios, $r_\mathrm{spot}$ was varied to find the detection limit in each scenario.

For the third host star, a K5V dwarf star was selected. Using an $T_\mathrm{eff} = 4100$\,K \citep[Table\,1][]{Sullivan2015}, the $R_\mathrm{*}$ and $M_*$ was extrapolated from the two polynomials presented by \citet{Boyajian2012} and are given in Table\,\ref{Tab.2}. We determine $u_1$ and $u_2$ from \citet{Claret2017}, for the K5V dwarf and are given in Table\,\ref{Tab.2}.

Because of the larger $R_\mathrm{*}$ the K5V simulations probed the detection limits of starspot anomalies for larger planets, 3-7\Rearth\ (Sub-Neptune to sub-Saturn class planets) in 0.5\Rearth\ increments. Due to the larger $R_\mathrm{*}$ and $R_\mathrm{p}$ the three values of $P$ were selected at 2,\,4\,\&\,6\,days with the corresponding values for $a$ calculated using Kepler's third law using a stellar mass of $0.65$\,\Msun\ (see Table\,\ref{Tab.2}) and are given in Table\,\ref{Tab.1}. With $T_\mathrm{eff} = 4100$\,K four temperatures were selected for the starspot 3700\,K, 3800\,K, 3900\,K and 4000\,K. Combined with the three observational wavelengths and four noise levels meant that a further 1296 scenarios were generated using an K5V dwarf host star.

\section{Simulation Results} \label{Sec:3}

For all 3888 different scenarios a total of 20573 simulated transits were generated. All 20573 simulated transit light curves will be made available at the CDS\footnote{\href{http://vizier.u-strasbg.fr/}{\tt http://vizier.u-strasbg.fr/}}. The number of simulated transits per scenario was dependent on the detection of $r_\mathrm{spot}$ and ranged from 2\,-\,40 simulations per scenario, where $r_\mathrm{spot}$ started at $0.0087R_\mathrm{*}$ and was increased in $0.0087R_\mathrm{*}$ steps up to $0.35R_\mathrm{*}$. The simulations were stopped when the starspot anomaly was deemed detectable. The starspot anomaly was considered ``detected'' in the simulated transits, if the characteristic ``blip'' was visible in the simulated light curve (e.g.,  Fig.\,\ref{fig:2}). This was accomplished by comparing the amplitude of the anomaly to the photometric precision (rms scatter) of the data. The amplitude of the starspot anomaly was measured by creating two transit light curves. The first had the centre of the starspot within the transit cord while the second light curve had the starspot latitude shifted so that it was no longer within the transit cord. Because both occulted and non-occulted starspots affect the shape of the light curve \citep{Silva2010} and in particular the transit depth \citep{Nikolov2013} it was important that the stellar disc of the ``spot-free" light curve model still contained a starspot of equal properties. To reduce the impact from foreshortening, the starspot was positioned so that only a gap of one pixel separated the starspot and transit cord boundaries. We compared the resulting light curves to determine the amount of systematic error the approximation would generate and found a mean divergence of $\approx 1$\,ppm, 60 times smaller than the TESS mission specified noise floor. Therefore, we conclude that this approximation is suitable for this work.

The light curve plots produced in this work show two synthetic light curves represented by a solid line and filled circles. The light curves represented by a solid line are the synthetic light curves containing the starspot anomaly and are without added noise. The light curves represented by filled circles are the synthetic light curves of the ``spot-free'' model and have added Gaussian noise. By comparing the two synthetic light curves it is possible to determine if the amplitude of the starspot anomaly is larger than the rms scatter.

To determine the most efficient set of detection conditions, tests were performed on the synthetic data. After the addition of noise to the synthetic light curve, \prism\ was fitted to the synthetic light curve in an attempt to recover the starspot parameters. The amplitude of the starspot anomaly was determined by the data points which described the starspot anomaly and the mean change in flux was calculated. The mean change in flux was then divided by the rms scatter of the light curve to give the amplitude of the starspot anomaly in units of rms ($\sigma$). Different amplitudes were considered from 1-$\sigma$ to 3-$\sigma$. In all cases where the amplitude was greater than 2.0-$\sigma$, the best fitting starspot parameters of the synthetic light curves were close to the original values and agreed within their 1-$\sigma$ parameter uncertainties. However, when the amplitude of the starspot anomaly was between 1.5-$\sigma$ and 2-$\sigma$ the 1-$\sigma$ uncertainty in the starspot parameters was non-physical. For example in one test the starspot radius was found to be $0.047R_\mathrm{*} \pm1.700R_\mathrm{*}$  indicating that the starspot covered the entire visible hemisphere of the stellar disc, and in another test the latitude of the starspot was found to be $0.17^\circ\pm97.39^\circ$, which indicated that a spot-free solution (i.e., the starspot was no longer on the transit cord) was an acceptable solution. These tests showed that when the amplitude of the starspot anomaly was between 1.5-$\sigma$ and 2-$\sigma$ the anomaly could be detected but it was not possible to constrain the starspot parameters, and so, confirm the presence of a starspot. Therefore, the optimal cut-off point for the amplitude of the starspot anomaly was set at the 2-$\sigma$ limit to differentiate between a detection and non-detection.

\subsection{M4V dwarf host star results} \label{Sec:3.1}

\begin{figure*} \includegraphics[width=0.48\textwidth,angle=0]{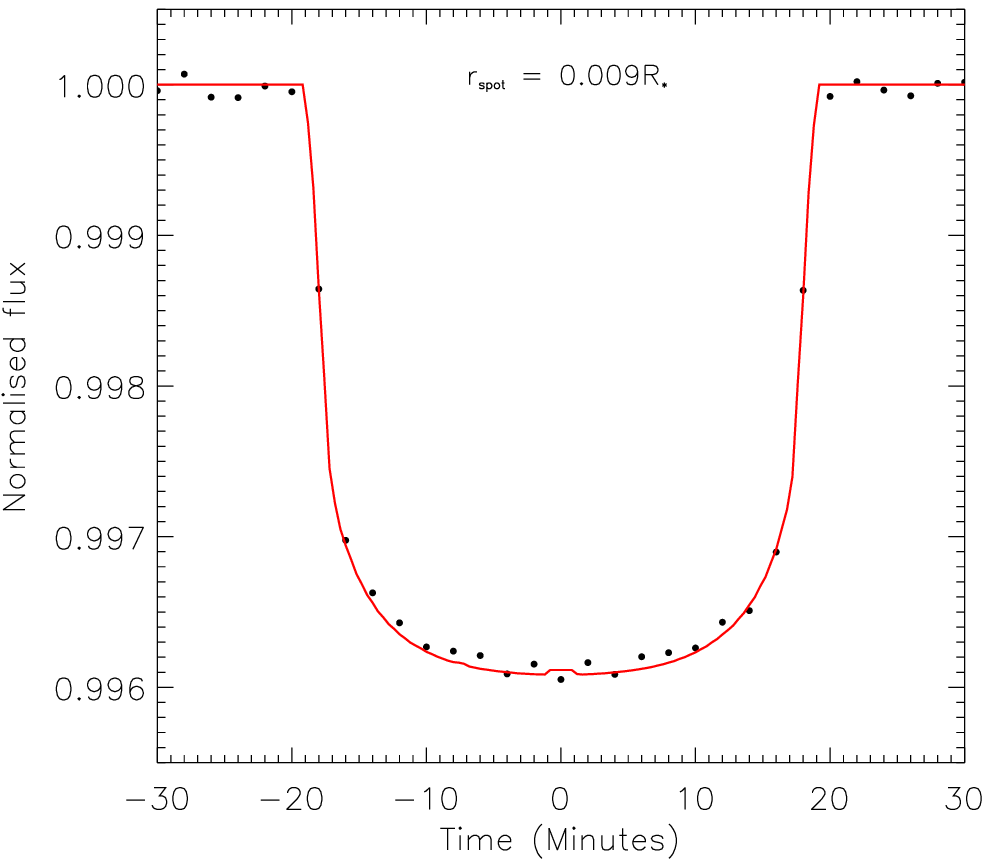} \includegraphics[width=0.48\textwidth,angle=0]{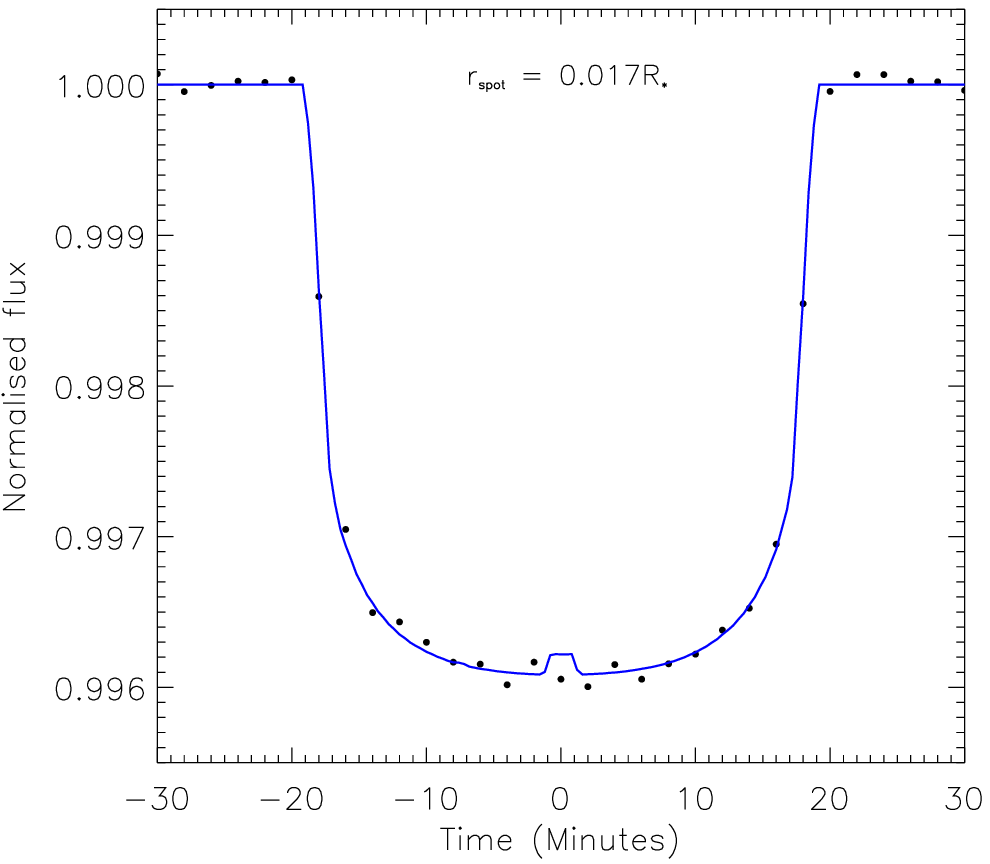}
\caption{\label{fig:3} Two simulated transit light curves generated by \prism. These were both generated using a 1.0\Rearth\ Earth-sized planet transiting a 3200\,K, 0.155\Rsun\ M4V dwarf star with $i =90.0^\circ$ and $P = 1$\,d. The starspot properties are: $\theta = 0^\circ$, $\phi = 90^\circ$, $r_\mathrm{spot} = 0.009R_\mathrm{*}$ (left), $r_\mathrm{spot} = 0.017R_\mathrm{*}$ (right) and $T_\mathrm{spot} = 3000$\,K. The observational wavelength and the rms scatter for the simulated transits were set at 600\,nm and 60\,ppm respectively. The starspot anomaly becomes visible over the simulated noise when $r_\mathrm{spot}$ is increased from $0.009R_\mathrm{*}$ to $0.017R_\mathrm{*}$. The solid lines represents the noise-free synthetic light curves containing the starspot anomaly, while, the filled circles represent the ``spot-free'' synthetic light curves (see Section\,\ref{Sec:3}) with added Gaussian noise.} \end{figure*}

A total of 6870 simulations were conducted for the 1296 scenarios using an M4V dwarf host star and the results are presented in Table\,\ref{Tab.M4_Results}. It was found that the starspot detection limit in TESS transit light curves of an M4V dwarf host star was $r_\mathrm{spot} = 0.045R_\mathrm{*}\pm 0.016R_\mathrm{*}$ ($4900\pm1700$\,km), where the uncertainty is the 1-$\sigma$ standard deviation from the 1296 scenarios of the M4V host star simulations. The smallest $r_\mathrm{spot}$ detected in the M4V simulations (under optimal parameter conditions) was $r_\mathrm{spot} = 0.017R_\mathrm{*}$ or $1900$\,km (see Fig.\,\ref{fig:3}) and was detected in 7 of the 1296 scenarios ($0.5$\,\%).

\begin{figure*} \includegraphics[width=1.0\textwidth,angle=0]{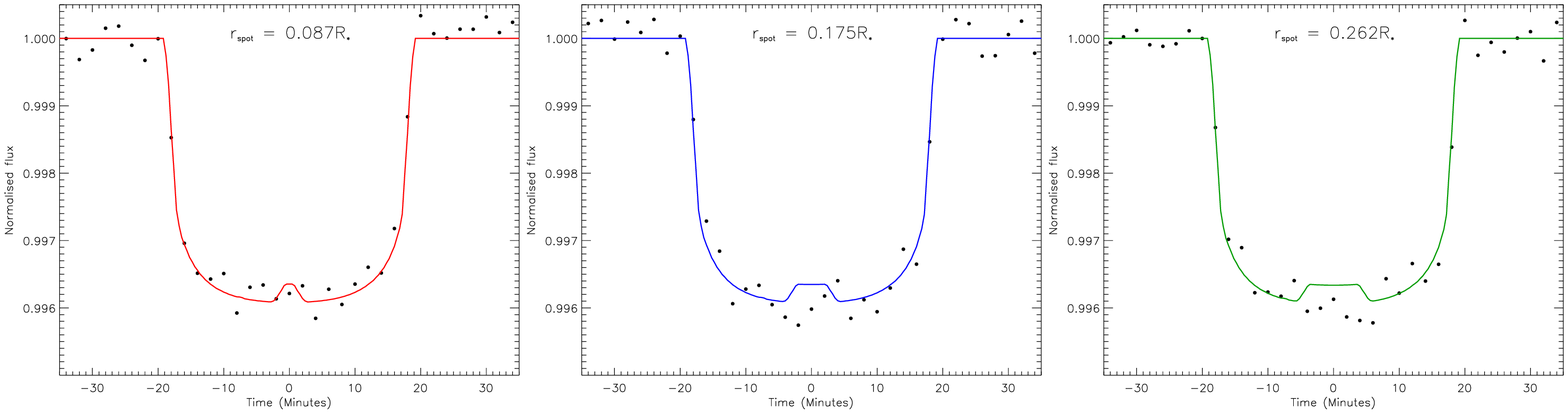}  \includegraphics[width=0.33\textwidth,angle=0]{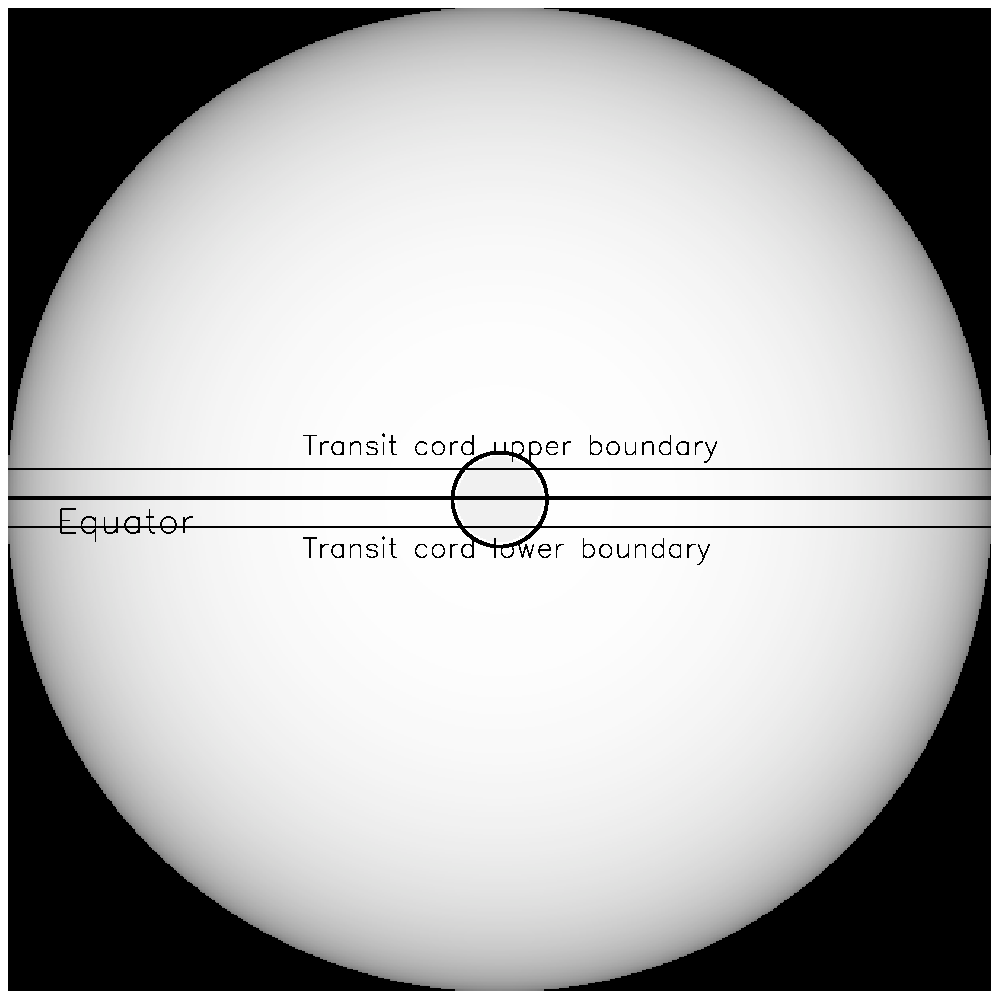} \includegraphics[width=0.33\textwidth,angle=0]{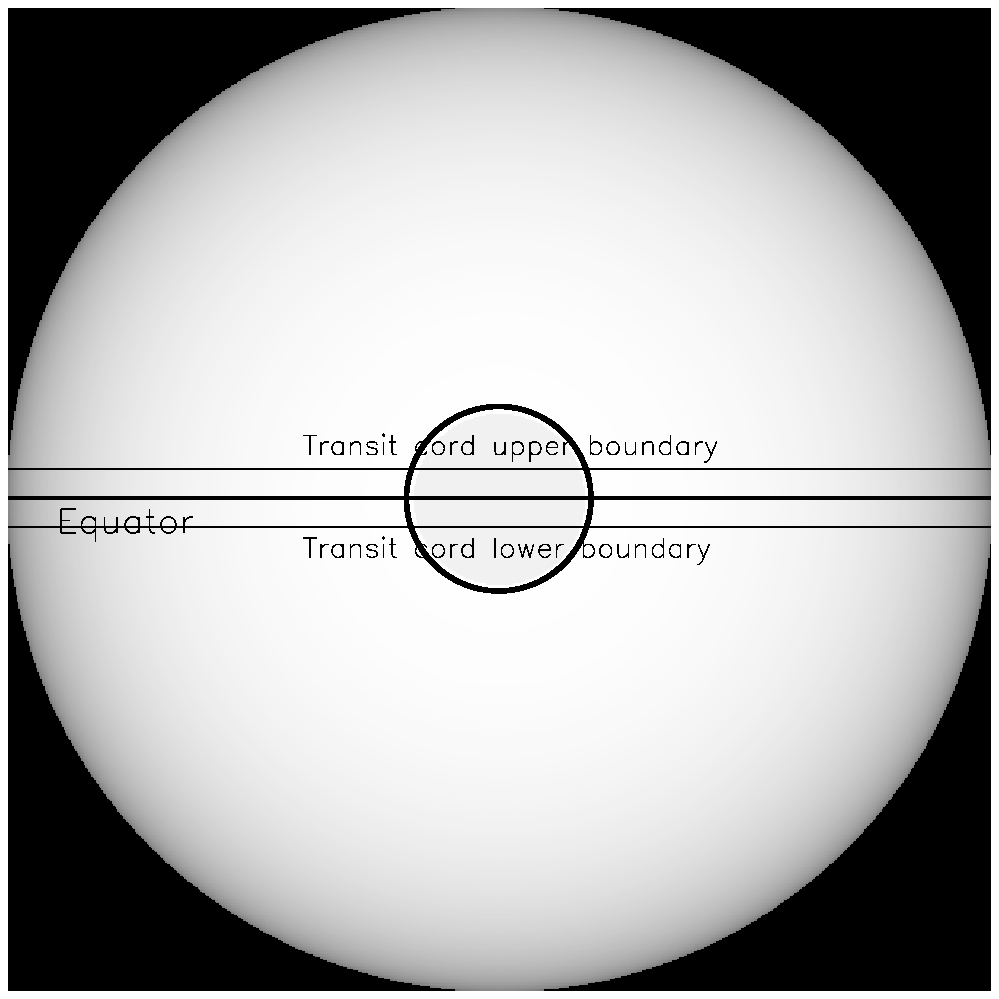} \includegraphics[width=0.33\textwidth,angle=0]{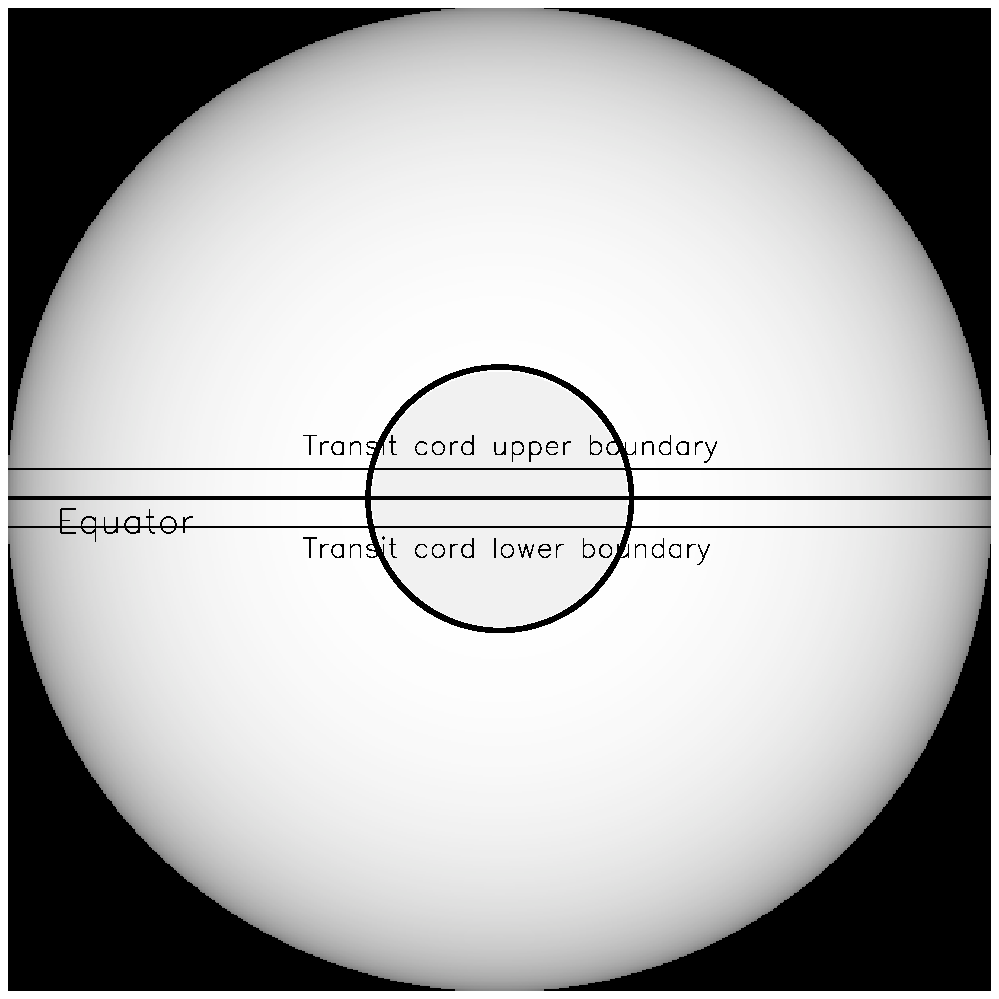} 
\caption{\label{fig:4} Three simulated transit light curves (top panels) and stellar disc outputs (bottom panels), generated by \prism. Light curves: The solid lines represents the noise-free synthetic light curves containing the starspot anomaly, while, the filled circles represent the ``spot-free'' synthetic light curves (see Section\,\ref{Sec:3}) with added Gaussian noise. Stellar discs: The central solid lines represents the stellar equator (assuming orbital alignment) and the upper and lower solid lines represent the upper and lower boundaries of the transit cords. Due to the high $\rho_\mathrm{spot}$, the starspot boundary has been highlighted with a black circular boarder. The observational wavelength and the rms scatter for the simulated transits were set at 1000\,nm and 200\,ppm respectively. These were generated using a 1.0\Rearth\ Earth-sized planet transiting a 3200\,K, 0.155\Rsun\ M4V dwarf star with $i =90.0^\circ$ and $P = 1$\,d. The starspot properties were set at: $\theta = 0^\circ$, $\phi = 90^\circ$ and $T_\mathrm{spot} = 3150$\,K ($\rho_\mathrm{spot} = 0.93$). The left panels have $r_\mathrm{spot} = 0.087R_\mathrm{*}$, the middle panels have $r_\mathrm{spot} = 0.175R_\mathrm{*}$ and the right panels have $r_\mathrm{spot} = 0.262R_\mathrm{*}$.} \end{figure*}

It was found that for three scenarios, it was not possible to resolve the starspot size (see Fig.\,\ref{fig:4}). These scenarios involved a 3150\,K starspot with a photometric precision of 200\,ppm, observed at 1000\,nm and with an 1\Rearth\ transiting planet. When examining the stellar disc output for the $0.093R_\mathrm{*}$ starspot in Fig.\,\ref{fig:4}, the area of the 2\,D projection of the starspot is larger than the planetary disc. Therefore, increasing $r_\mathrm{spot}$ further, would no longer increase the amplitude of the starspot anomaly, because it would not affect the amount of flux blocked by the planet. However, it will increase the duration, due to the fact, that the planet will spend longer occulting the starspot. This can be seen when examining the three simulated light curves in Fig.\,\ref{fig:4}. Therefore, for these scenarios, the amplitude of the starspot anomaly never became larger than the rms scatter of the synthetic data. 

Hence, the simulations show that it is not possible to observe a 3150\,K starspot on the surface of a 3200\,K M4V dwarf, using a 1.0\Rearth\ Earth-sized planet, when observed at the reddest edge of the $T$ passband (i.e., 1000\,nm) with an rms scatter of 200\,ppm. However, by either increasing the planetary radius ($+0.25$\Rearth), reducing the rms scatter ($-50$\,ppm) or shifting to a bluer wavelength, it would be possible to detect these starspots down to $0.035R_\mathrm{*}$ ($3800$\,km) (see Table\,\ref{Tab.M4_Results}).

\subsection{M1V dwarf host star results} \label{Sec:3.2}

\begin{figure*} \includegraphics[width=0.48\textwidth,angle=0]{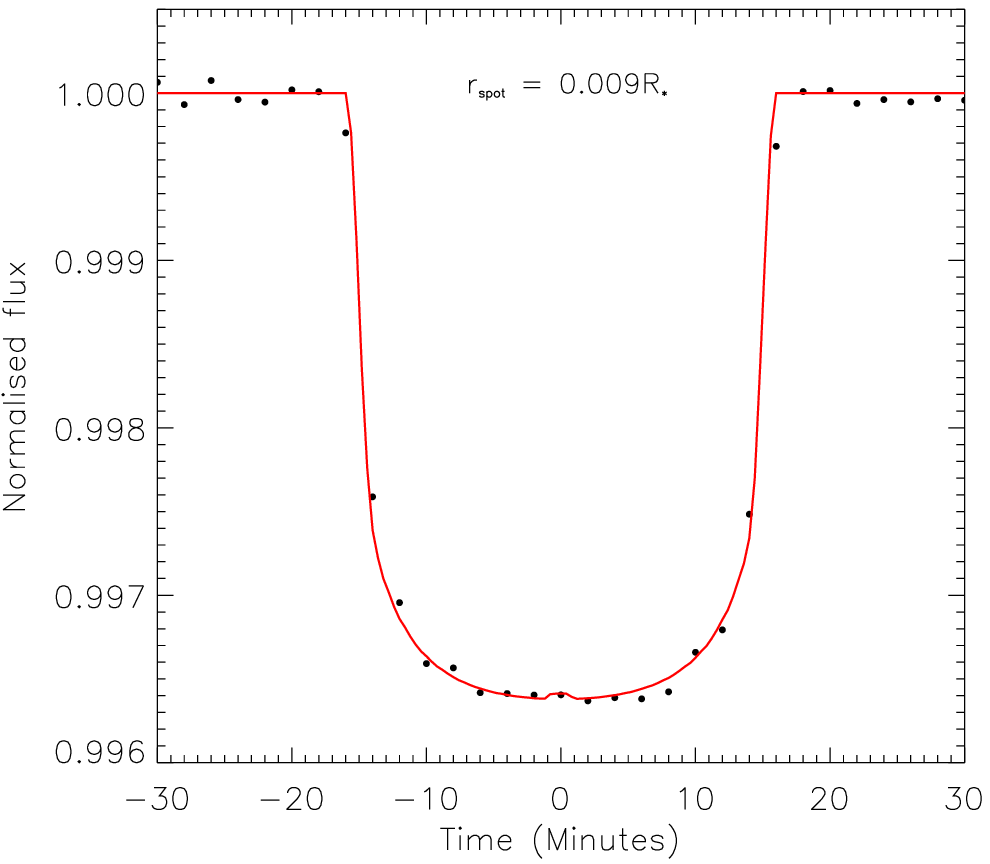} \includegraphics[width=0.48\textwidth,angle=0]{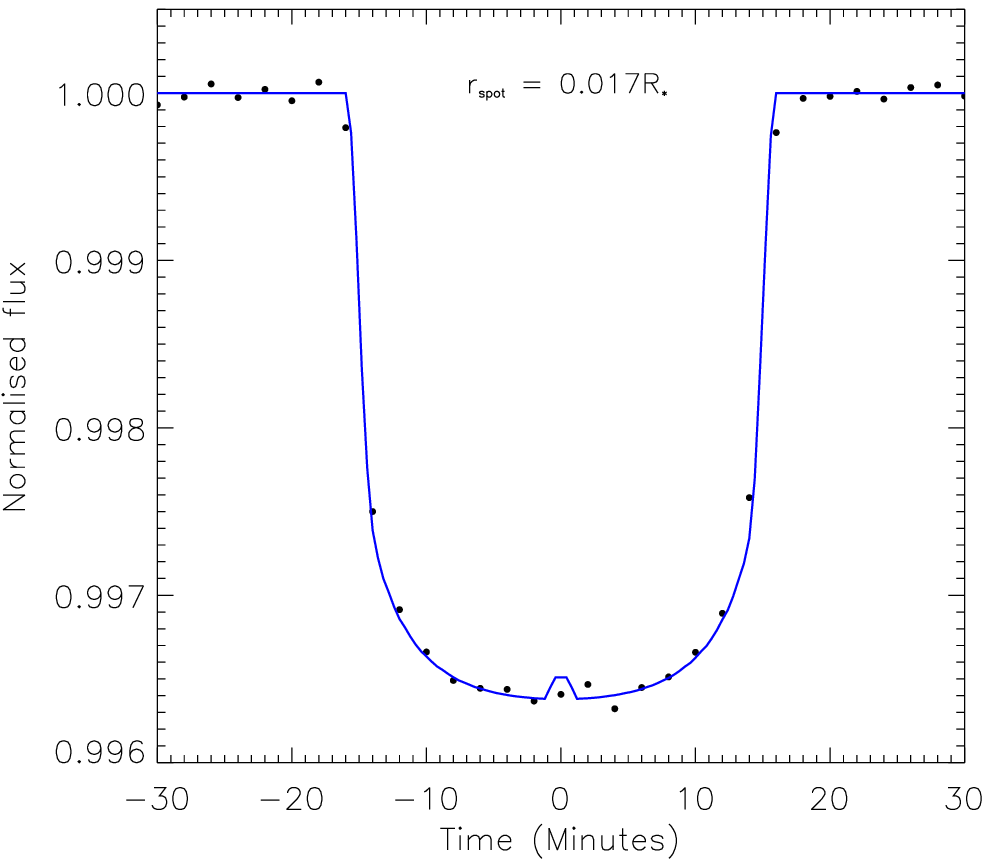}
\caption{\label{fig:5} Two simulated transit light curves generated by \prism. These were both generated using a 2.75\Rearth\ sub-Neptune planet transiting a 3700\,K, 0.493\Rsun\ M1V dwarf star with $i =90.0^\circ$ and $P = 3$\,d. The starspot properties are: $\theta = 0^\circ$, $\phi = 90^\circ$, $r_\mathrm{spot} = 0.009R_\mathrm{*}$ (left), $r_\mathrm{spot} = 0.017R_\mathrm{*}$ (right) and $T_\mathrm{spot} = 3400$\,K. The observational wavelength and the rms scatter for the simulated transits were set at 785\,nm and 60\,ppm respectively. The starspot anomaly becomes visible over the simulated noise when $r_\mathrm{spot}$ is increased from $0.009R_\mathrm{*}$ to $0.017R_\mathrm{*}$. The solid lines represents the noise-free synthetic light curves containing the starspot anomaly, while, the filled circles represent the ``spot-free'' synthetic light curves (see Section\,\ref{Sec:3}) with added Gaussian noise.} \end{figure*}

A total of 7769 simulations were conducted for the 1296 scenarios using an M1V dwarf host star and the results are presented in Table\,\ref{Tab.M1_Results}. It was found that the starspot detection limit in TESS transit light curves of an M1V dwarf host star was $r_\mathrm{spot} = 0.040R_\mathrm{*}\pm 0.017R_\mathrm{*}$ ($13800\pm6000$\,km). The smallest $r_\mathrm{spot}$ detected in the M1V simulations (under optimal parameter conditions) was $r_\mathrm{spot} = 0.017R_\mathrm{*}$ or $6000$\,km (see Fig.\,\ref{fig:5}) and was detected in 85 of the 1296 scenarios ($6.6$\,\%).

To allow for a direct comparison between the starspots on the M1V and M4V dwarf stars requires the M1V dwarf star, $r_\mathrm{spot}$ to be multiplied by $\left(\frac{R_\mathrm{M1V}}{R_\mathrm{M4V}}\right) \approx 3.2$. From this it can be seen that while the smallest starspot detected in the M1V dwarf simulations had $r_\mathrm{spot} = 0.017R_\mathrm{*}$, this is the equivalent to $r_\mathrm{spot} \approx 0.056R_\mathrm{*}$ for a M4V dwarf star.

Similar to the M4V host star results, the M1V host star results show that it is not possible to detect the hottest starspots when small planets (e.g., $<2.75$\Rearth) are transiting the star, if the photometric precision is $>100$\,ppm.

\subsection{K5V dwarf host star results} \label{Sec:3.3}

\begin{figure*} \includegraphics[width=0.48\textwidth,angle=0]{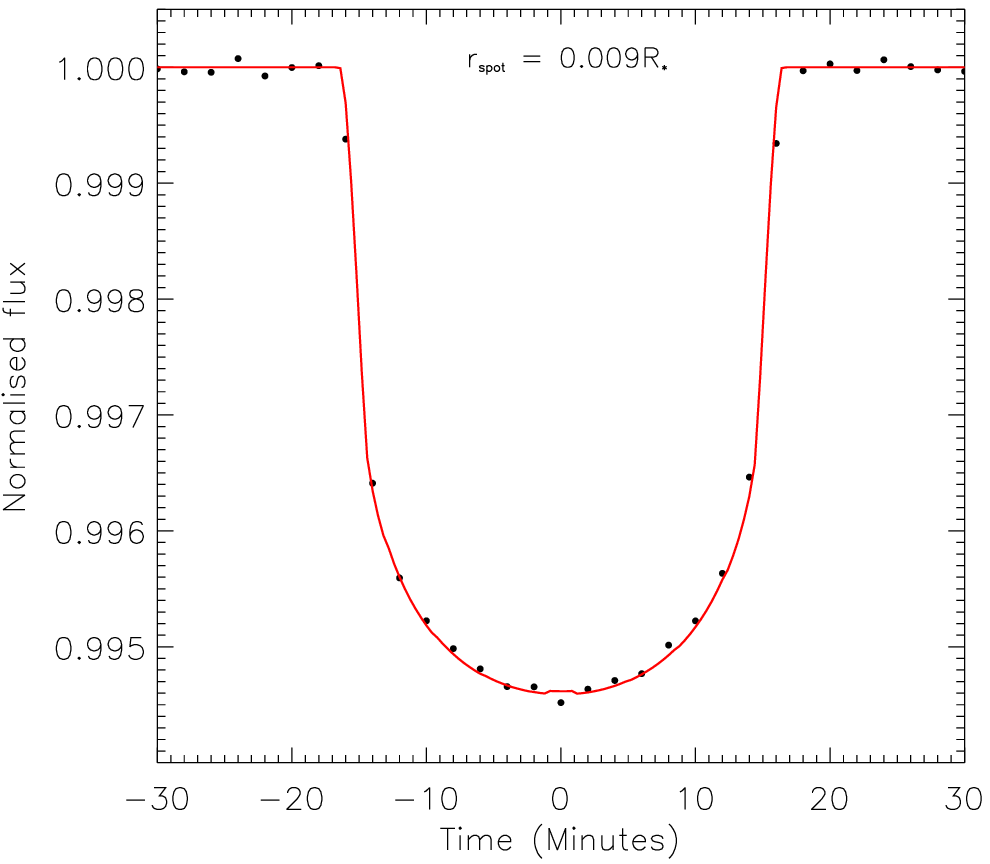} \includegraphics[width=0.48\textwidth,angle=0]{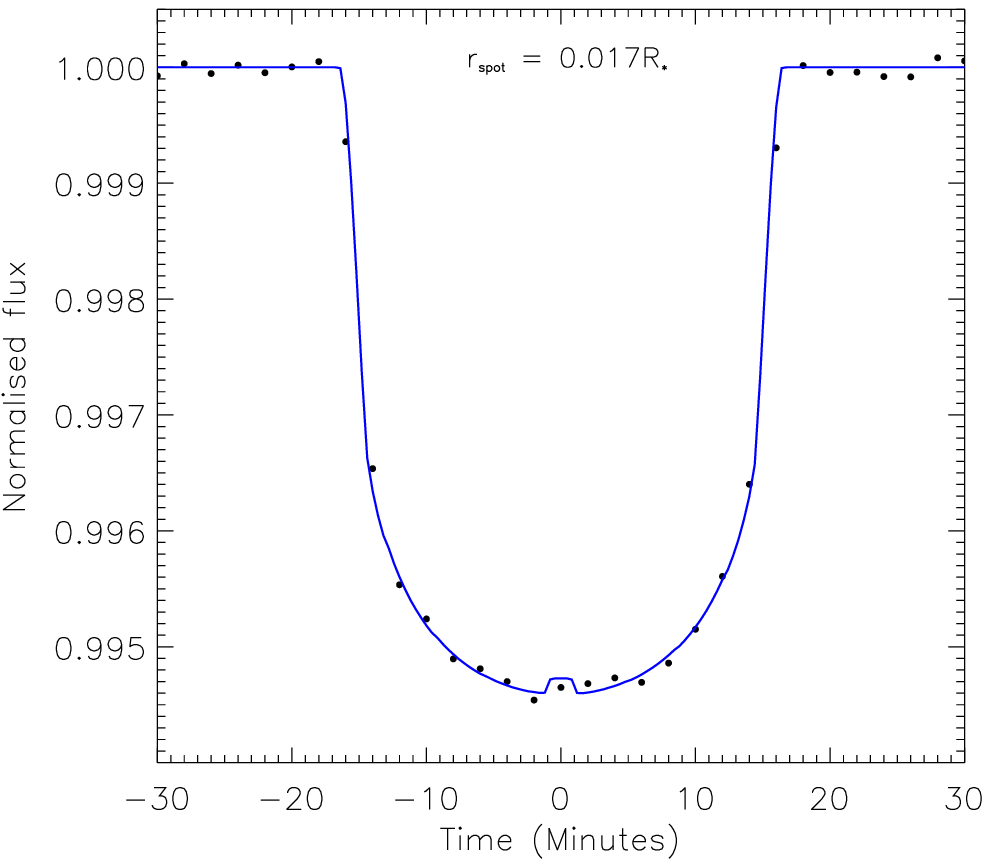}
\caption{\label{fig:6} Two simulated transit light curves generated by \prism. These were both generated using a 4.5\Rearth\ Neptune-sized planet transiting a 4100\,K, 0.623\Rsun\ K5V dwarf star with $i =90.0^\circ$ and $P = 4$\,d. The starspot properties are: $\theta = 0^\circ$, $\phi = 90^\circ$, $r_\mathrm{spot} = 0.009R_\mathrm{*}$ (red), $r_\mathrm{spot} = 0.017R_\mathrm{*}$ (blue) and $T_\mathrm{spot} = 3700$\,K. The observational wavelength and the rms scatter for the simulated transits were set at 785\,nm and 60\,ppm respectively. The starspot anomaly becomes visible over the simulated noise when $r_\mathrm{spot}$ is increased from $0.009R_\mathrm{*}$ to $0.017R_\mathrm{*}$. The solid lines represents the noise-free synthetic light curves containing the starspot anomaly, while, the filled circles represent the ``spot-free'' synthetic light curves (see Section\,\ref{Sec:3}) with added Gaussian noise.} \end{figure*}

A total of 5934 simulations were conducted for the 1296 scenarios using a K5V dwarf host star and the results are presented in Table\,\ref{Tab.K5_Results}. It was found that the starspot detection limit in TESS transit light curves of a K5V dwarf host star was $r_\mathrm{spot} = 0.038R_\mathrm{*}\pm 0.016R_\mathrm{*}$ ($15900\pm6800$\,km). The smallest $r_\mathrm{spot}$ detected in the K5V simulations (under optimal parameter conditions) was $r_\mathrm{spot} = 0.017R_\mathrm{*}$ or $7600$\,km (see Fig.\,\ref{fig:6}) and was detected in 102 of the 1296 scenarios ($7.9$\,\%).

We use the same process described in Section\,\ref{Sec:3.2} to compare the starspot sizes on the K5V and M4V dwarf stars. We find that for the K5V dwarf star, $r_\mathrm{spot}$ should be multiplied by $\left(\frac{0.623}{0.155}\right) \approx 4.0$. The smallest starspot detections for the K5V host star simulations were when $r_\mathrm{spot} = 0.017R_\mathrm{*}$, which if the starspots were on a M4V dwarf instead, would equate to a detection with $r_\mathrm{spot} \approx 0.07R_\mathrm{*}$.

It was found that for only 15 (1.2\,\%) scenarios it was not possible to resolve the starspot. These scenarios followed the same pattern as found with the previous host stars. The scenarios involving a high temperature starspot, small planet and observed at redder wavelengths with low photometric precision, tended to prevent the detection of the starspot.

\begin{figure*} \includegraphics[width=1.0\textwidth,angle=0]{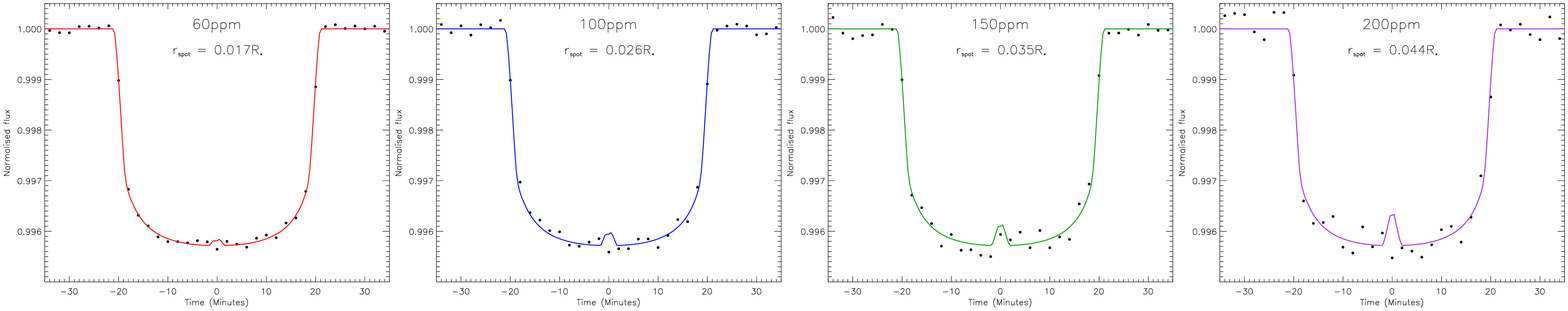} 
\caption{\label{fig:7} Four simulated transit light curves, generated by \prism. The observational wavelength of the simulated transits was 785\,nm. These were generated using a 3.0\Rearth\ sub-Neptune planet transiting a 3700\,K, 0.493\Rsun\ M1V dwarf star with $i =90.0^\circ$ and $P = 2$\,d. The starspot properties were set at: $\theta = 0^\circ$, $\phi = 90^\circ$ and $T_\mathrm{spot} = 3475$\,K. The rms scatter of the light curves from left to right are 60\,ppm, 100\,ppm, 150\,ppm and 200\,ppm. Each light curve shows the smallest detectable starspot size for the scenarios, $0.017R_\mathrm{*}$ (red), $0.026R_\mathrm{*}$ (blue), $0.035R_\mathrm{*}$ (green) and $0.044R_\mathrm{*}$ (purple), which increases with increasing noise. The solid lines represents the noise-free synthetic light curves containing the starspot anomaly, while the filled circles represent the ``spot-free'' synthetic light curves (see Section\,\ref{Sec:3}) with added Gaussian noise.} \end{figure*}

\begin{figure*} \includegraphics[width=1.0\textwidth,angle=0]{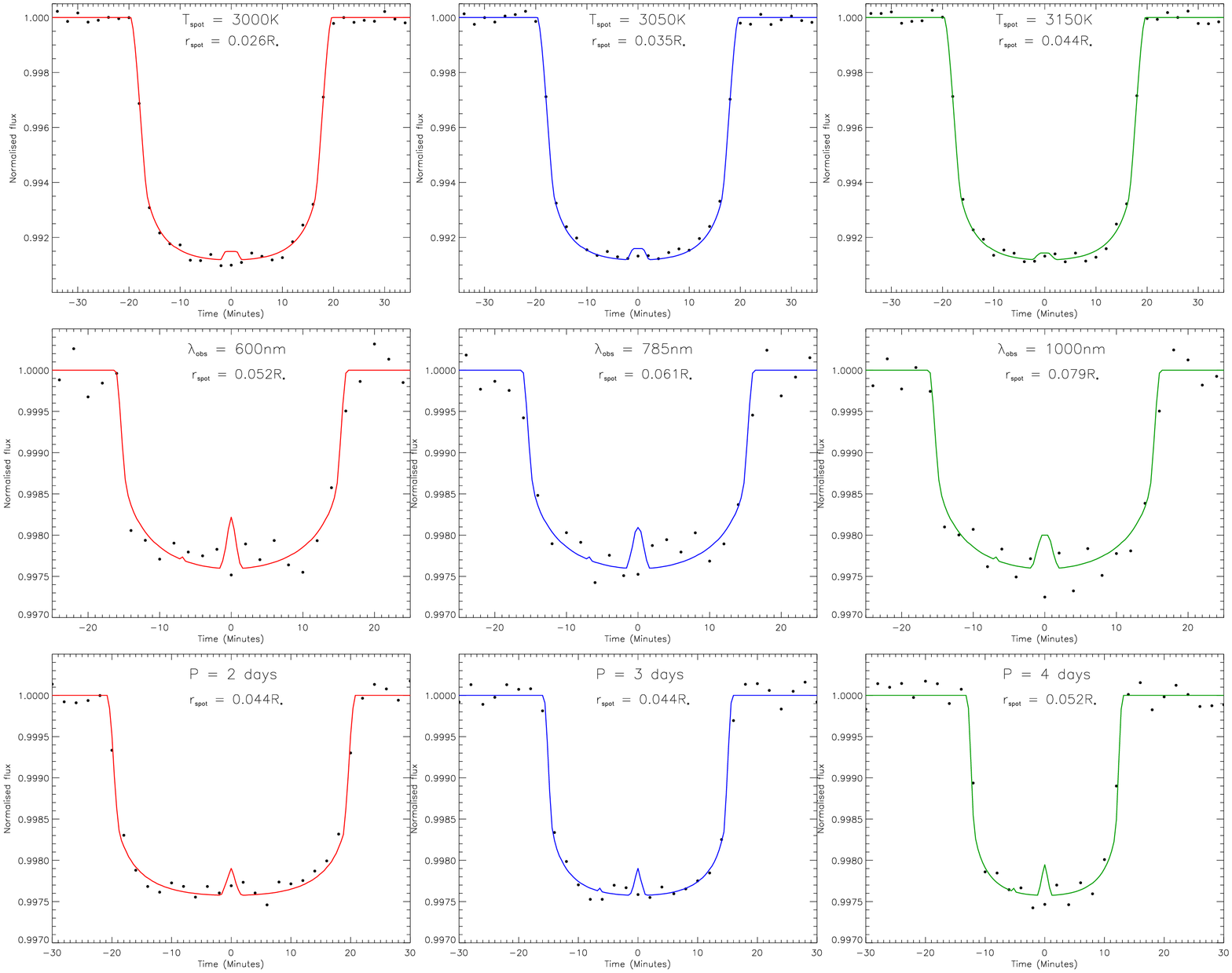}
\caption{\label{fig:8} Nine simulated transit light curves, generated by \prism, showing the smallest detected starspot for three trends, $T_\mathrm{eff}$: top row; $\lambda_\mathrm{obs}$: middle row; $P$: bottom row. ({\it Top row}) simulations of a 1.5\Rearth\ super-Earth planet transiting a 3200\,K, 0.155\Rsun\ M4V dwarf star with $i =90.0^\circ$, $P = 1$\,d, $\lambda_\mathrm{obs}=600$\,nm and $T_\mathrm{spot}=3000$\,K, $3050$\,K and $3100$\,K (left to right). With rms scatter $150\pm7.5$\,ppm. The smallest detected $r_\mathrm{spot}$ are: $r_\mathrm{spot}=0.026R_\mathrm{*}$ (red), $r_\mathrm{spot}=0.035R_\mathrm{*}$ (blue), $r_\mathrm{spot}=0.044R_\mathrm{*}$ (green). ({\it Middle row}) simulations of a 3.0\Rearth\ Neptune size planet transiting a 4100\,K, 0.623\Rsun\ K5V dwarf star with $i =90.0^\circ$, $P = 4$\,d, $T_\mathrm{spot}=3900$\,K and $\lambda_\mathrm{obs}=600$\,nm, $785$\,nm and $1000$\,nm (left to right). With rms scatter $200\pm10$\,ppm. The smallest detected $r_\mathrm{spot}$ values are: $r_\mathrm{spot}=0.052R_\mathrm{*}$ (red), $r_\mathrm{spot}=0.061R_\mathrm{*}$ (blue), $r_\mathrm{spot}=0.079R_\mathrm{*}$ (green). ({\it Bottom row}) simulations of a 2.25\Rearth\ sub-Neptune planet transiting a 3700\,K, 0.493\Rsun\ M1V dwarf star with $i =90.0^\circ$, $T_\mathrm{spot}=3550$\,K, $\lambda_\mathrm{obs}=1000$\,nm and $P = 2$\,d, $3$\,d and $4$\,d (left to right). With rms scatter $100\pm5$\,ppm. The smallest detected $r_\mathrm{spot}$ are: $r_\mathrm{spot}=0.044R_\mathrm{*}$ (red), $r_\mathrm{spot}=0.044R_\mathrm{*}$ (blue), $r_\mathrm{spot}=0.052R_\mathrm{*}$ (green). The solid lines represents the noise-free synthetic light curves containing the starspot anomaly, while, the filled circles represent the ``spot-free'' synthetic light curves (see Section\,\ref{Sec:3}) with added Gaussian noise.} \end{figure*}

\subsection{Expected trends} \label{Sec:3.4}

By examining the results from the 20573 simulated transits (see Tables\,\ref{Tab.M4_Results},\,\ref{Tab.M1_Results},\,\ref{Tab.K5_Results}), known trends (described in Section\,\ref{Sec:2.2}) in the starspot detection limits are seen.

The simulations confirmed that by increasing the photometric precision (decrease in noise) in the observations, allowed smaller starspots to be detected (e.g., Fig.\,\ref{fig:7}). The simulations also, confirm that the starspot detection limit is dependent on $T_\mathrm{spot}$, the observational wavelength and the orbital period (e.g., Fig.\,\ref{fig:8}). It is easier to detect a small cool starspot than a small hot starspot (top row\,Fig.\,\ref{fig:8}). Likewise, a shorter orbital period allows the detection of smaller starspots (bottom row\,Fig.\,\ref{fig:8}). This is due to a longer transit duration and by association, a longer starspot anomaly duration. Thereby increasing the number of data points describing both the transit and the starspot anomaly, due to the fixed 2\,min cadence of TESS data.

The simulations help support the predicted / known trends given in Section\,\ref{Sec:2.2} and highlights which parameters the detection limit is strongly or weakly, dependent on. Examining Fig.\,\ref{fig:8}, it can seen that by doubling the orbital period ($P$), the minimum detected $r_\mathrm{spot}$ increases by 20\,\%, while shifting the observed wavelength from 600\,nm to 1000\,nm, the minimum detected $r_\mathrm{spot}$ increases by 50\,\%. However increasing $T_\mathrm{spot}$ by 100\,K increases the minimum detected $r_\mathrm{spot}$ by 67\,\%\footnote{For a 3200\,K M4V dwarf star, observed at 600\,nm.}, indicating that the detection limit is primarily influenced by $T_\mathrm{spot}$ and $\lambda_\mathrm{obs}$ -- which is expected -- due to the amplitude of a starspot anomaly being directly proportional to $\rho_\mathrm{spot}$ \citep[see][]{Jeremy2012}, which in turn is proportional to $T_\mathrm{spot}$ and $\lambda_\mathrm{obs}$ (see Eq.\,\ref{eq.1}).

The amplitude of the starspot anomaly is proportional to $\left(\frac{r_\mathrm{spot}}{R_\mathrm{p}}\right)^2$ when $r_\mathrm{spot} \le R_\mathrm{p}$.  Therefore, smaller planets can detect smaller starspots. The results from the simulations agree with this when dealing with cool starspots. However, the simulations depict a different trend when dealing with the hotter starspots, in that, larger planets are able to detect smaller starspots. For example in Table\,\ref{Tab.M4_Results} we can see that a 1.25\Rearth\ planet transiting a 3200\,K, 0.155\Rsun\ M4V dwarf star with $i =90.0^\circ$, $P = 2$\,d, $\lambda_\mathrm{obs}=785$\,nm and $T_\mathrm{spot}=3150$\,K, with rms scatter $150\pm7.5$\,ppm, the smallest detected $r_\mathrm{spot}$ is $r_\mathrm{spot}=0.070R_\mathrm{*}$ While for a 1.5\Rearth\ and 1.75\Rearth\ planet the smallest detected $r_\mathrm{spot}$ becomes $r_\mathrm{spot}=0.061R_\mathrm{*}$ and $r_\mathrm{spot}=0.052R_\mathrm{*}$ respectively.

\subsection{Unexpected trend} \label{Sec:3.5}

Table\,\ref{Tab.3} shows the comparison between $R_\mathrm{p}$ and the statistical mean of the starspot detection limit ($\bar{r}_\mathrm{spot}$), including its 1-$\sigma$ uncertainty. At first glance, Table\,\ref{Tab.3} seems to be in contradiction with the statement ``small planets are needed to detect small starspots''. However,  looking at the top row of Table\,\ref{Tab.3}, we see that when the 1-$\sigma$ uncertainty in $\bar{r}_\mathrm{spot}$ is taken into account, we see that the smallest planet (e.g., $R_\mathrm{p} = 1$\,\Rearth) can detect the smallest starspot (e.g., $\bar{r}_\mathrm{spot} = 0.018R_\mathrm{*}$). The results in Table\,\ref{Tab.3} also indicate that the range of values for $\bar{r}_\mathrm{spot}$, reduce with increasing $R_\mathrm{p}$. 

When examining tables\,\ref{Tab.M4_Results},\,\ref{Tab.M1_Results},\,\ref{Tab.K5_Results}, we can see an unexpected trend emerging: starspots which are small and hot can be detected by large planets, however, they can't be detected by smaller planets. This unexpected and counter-intuitive trend, however, is a manifestation, from using a fixed cadence combined with the methodology used to detect starspot anomalies in transit light curves.

\begin{table*} \centering
\caption{\label{Tab.3} Comparison between $R_\mathrm{p}$ and $\bar{r}_\mathrm{spot}$ for the M4V, M1V and K5V host stars.}
\setlength{\tabcolsep}{5pt} \vspace{-5pt}
\begin{tabular}{cccccc} 
\hline\hline
 \multicolumn{2}{c}{M4V}  &  \multicolumn{2}{c}{M1V}  &   \multicolumn{2}{c}{K5V}     \\    
   $R_\mathrm{p}$\,(\Rearth) & $\bar{r}_\mathrm{spot}$\,($R_\mathrm{*}$) &  $R_\mathrm{p}$\,(\Rearth) & $\bar{r}_\mathrm{spot}$\,($R_\mathrm{*}$) & $R_\mathrm{p}$\,(\Rearth) & $\bar{r}_\mathrm{spot}$\,($R_\mathrm{*}$)   \\   
\hline
  1.00 &  $0.042\pm0.024$ &  2.00 & $0.045\pm0.028$ &  3.0 & $0.044\pm0.031$  \\ 
  1.25 &  $0.038\pm0.014$ &  2.25 & $0.044\pm0.023$ &  3.5 & $0.040\pm0.023$  \\ 
  1.50 &  $0.038\pm0.012$ &  2.50 & $0.042\pm0.021$ &  4.0 & $0.037\pm0.017$  \\ 
  1.75 &  $0.038\pm0.010$ &  2.75 & $0.042\pm0.021$ &  4.5 & $0.035\pm0.014$  \\ 
  2.00 &  $0.040\pm0.010$ &  3.00 & $0.040\pm0.019$ &  5.0 & $0.035\pm0.012$  \\ 
  2.25 &  $0.044\pm0.010$ &  3.25 & $0.038\pm0.016$ &  5.5 & $0.035\pm0.012$  \\ 
  2.50 &  $0.045\pm0.010$ &  3.50 & $0.037\pm0.014$ &  6.0 & $0.035\pm0.010$  \\ 
  2.75 &  $0.051\pm0.014$ &  3.75 & $0.037\pm0.014$ &  6.5 & $0.035\pm0.010$  \\ 
  3.00 &  $0.054\pm0.014$ &  4.00 & $0.037\pm0.012$ &  7.0 & $0.035\pm0.010$  \\ 
\hline \end{tabular}
\end{table*}

\begin{figure} \includegraphics[width=0.48\textwidth,angle=0]{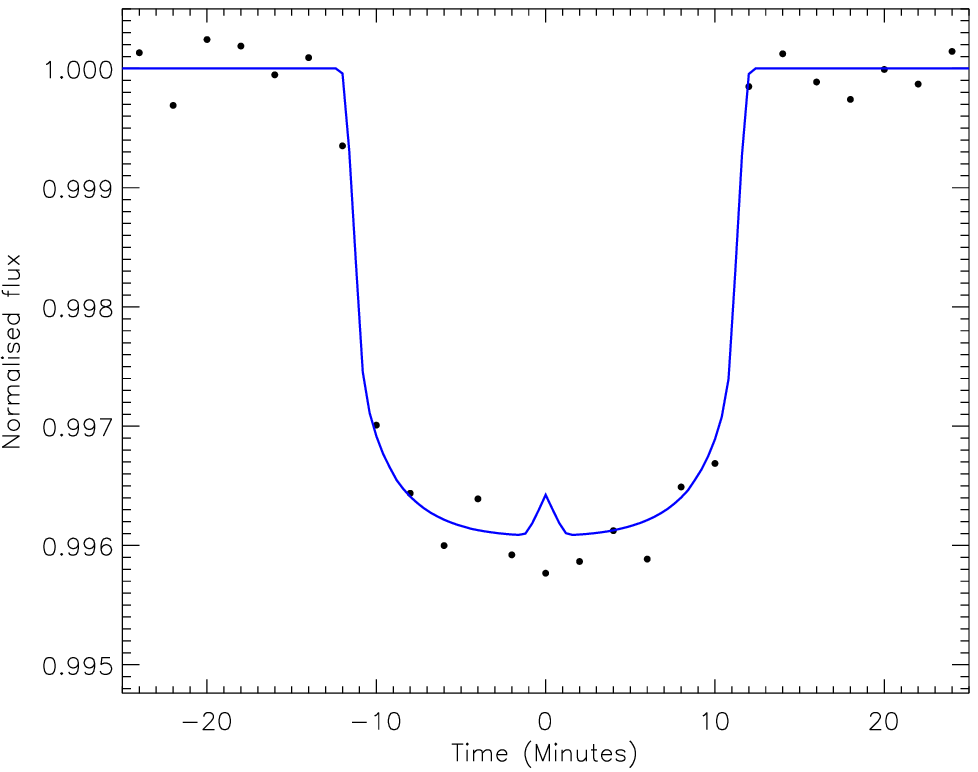} 
\caption{\label{fig:9} Example of a single data point describing the apex of a starspot anomaly. Using a 1.0\Rearth\ Earth-sized planet transiting a 3200\,K, 0.155\Rsun\ M4V dwarf star with $i =90.0^\circ$ and $P = 2$\,d. The starspot properties are: $\theta = 0^\circ$, $\phi = 90^\circ$, $r_\mathrm{spot}=0.061R_\mathrm{*}$ and $T_\mathrm{spot} = 3150$\,K. The observational wavelength and the rms scatter for the simulated data were set at 785\,nm and $200\pm10$\,ppm respectively. The solid line represents the noise-free synthetic light curve containing the starspot anomaly, while, the filled circles represent the ``spot-free'' synthetic light curve (see Section\,\ref{Sec:3}) with added Gaussian noise.} \end{figure}

\begin{figure} \includegraphics[width=0.48\textwidth,angle=0]{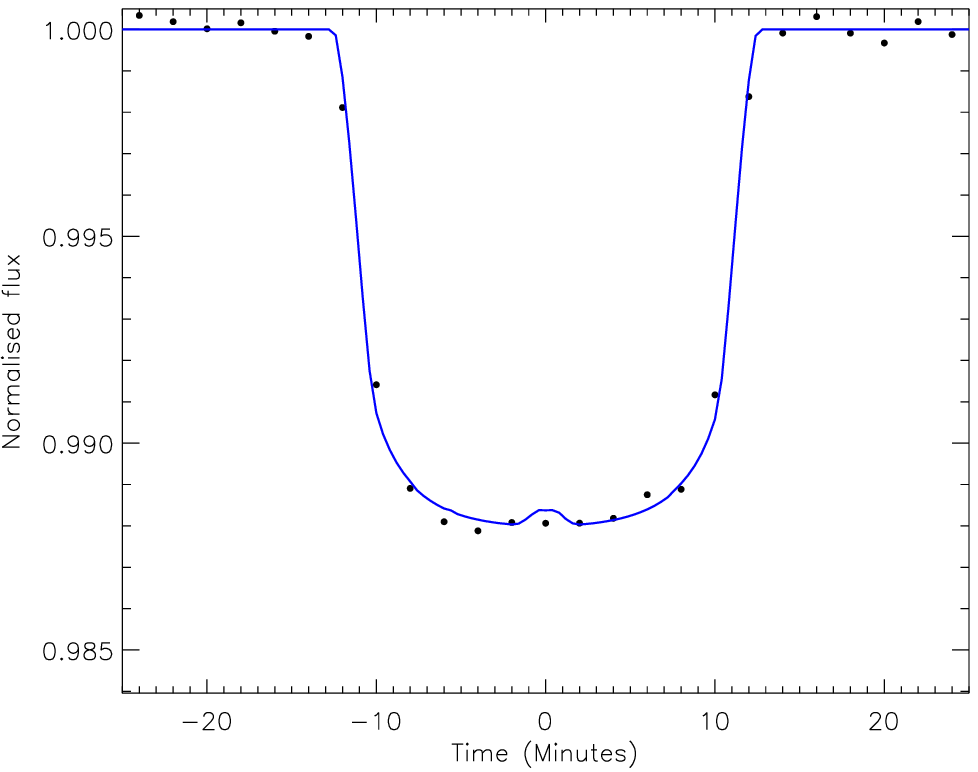} 
\caption{\label{fig:10} Example of multiple data points describing the plateau of a starspot anomaly. Using a 1.75\Rearth\ super-Earth planet transiting a 3200\,K, 0.155\Rsun\ M4V dwarf star with $i =90.0^\circ$ and $P = 2$\,d. The starspot properties are: $\theta = 0^\circ$, $\phi = 90^\circ$, $r_\mathrm{spot}=0.061R_\mathrm{*}$ and $T_\mathrm{spot} = 3150$\,K. The observational wavelength and the rms scatter for the simulated data were set at 785\,nm and $200\pm10$\,ppm respectively. The solid line represents the noise-free synthetic light curve containing the starspot anomaly, while, the filled circles represent the ``spot-free'' synthetic light curve (see Section\,\ref{Sec:3}) with added Gaussian noise.} \end{figure}

Increasing $R_\mathrm{p}$ will not only increase the transit duration, it will also increase the duration of the starspot anomaly. This is because a larger planetary disc will spend longer occulting the starspot. For a fixed cadence (2\,mins for TESS) this will equate to an increase in the number of data points which will describe the starspot anomaly. As mentioned in Section\,\ref{Sec:3} the mean flux of the data points within the starspot anomaly was used to determine if the starspot had been detected. This is in principle the same method used to detect starspot anomalies in observed light curves. In the case when a single data point describes the apex of a starspot anomaly (i.e., the 2\,D projection of $r_\mathrm{spot}$ equals $R_\mathrm{p}$; Fig.\,\ref{fig:9}) the data points which lie within the ingress and egress of the starspot anomaly will reduce the mean flux. For a starspot anomaly that’s described by multiple data points which form a plateau (i.e., the 2\,D projection of $r_\mathrm{spot}$ is either larger or smaller than $R_\mathrm{p}$; Fig.\,\ref{fig:10}) the mean flux will increase due to the data points lying on the starspot ingress and egress, having a weaker impact. Because of this, a starspot with an apex will require a larger amplitude to be detected. This is correct as a single data point should require a larger amplitude (compared to multiple data points) to allow a confirmed starspot detection. However, this creates an unavoidable detection bias.

Fig.\,\ref{fig:9}\,\&\,Fig.\,\ref{fig:10} depict two identical scenarios except that they have two different $R_\mathrm{p}$, $R_\mathrm{p}=1.0$\,\Rearth\ (Fig.\,\ref{fig:9}) and $R_\mathrm{p}=1.75$\,\Rearth\ (Fig.\,\ref{fig:10}). The starspot ($r_\mathrm{spot}=0.061R_\mathrm{*}$) is not detected in Fig.\,\ref{fig:9}, however, by increasing the duration of the starspot anomaly, and therefore, the number of data points which describe it, the starspot is then detected (Fig.\,\ref{fig:10}).

The mean amplitude of the starspot anomalies are 205\,ppm (Fig.\,\ref{fig:9}) and 394\,ppm (Fig.\,\ref{fig:10}). This shows how by increasing the duration of starspot anomaly, can allow for its detection and explains why large planets are needed to detect starspots which are both small and hot in TESS transit light curves.

\subsection{Comparing the change in flux between the planet and starspot} \label{Sec:3.6}

The values of $R_\mathrm{p}$ and $\bar{r}_\mathrm{spot}$ given in Table\,\ref{Tab.3}, while are valuable, don't allow for a direct comparison between the three host star spectral types. To allow a direct comparison, $R_\mathrm{p}$ and $\bar{r}_\mathrm{spot}$ need to be converted into change in flux ($\Delta F$). $R_\mathrm{p}$ can be converted to $\Delta F_\mathrm{p}$ using $k^2$ where:

\begin{equation}\label{Eq.5}
\begin{split}
\Delta F_\mathrm{p} = & \ k^2 \\
= & \ \left(\frac{R_\mathrm{p}}{R_\mathrm{*}}\right)^2\ ,
\end{split}
\end{equation}

\noindent which gives the area of the planetary disc in units of a uniform stellar disc. Because the simulations used in this work examined starspot occultations when the starspot was at the centre of the stellar disc, we can use this approximation of $\Delta F_\mathrm{p}$ in the absence of limb darkening.

The surface area of the starspot is the solid angle, $\Omega$ and using $r_\mathrm{spot}$ as the angular radius of the starspot, is given by:

\begin{equation}
\Omega = 2\pi R^2_\mathrm{*}\left(1 - \cos r_\mathrm{spot}\right)\ ,
\end{equation}

\noindent which in units of the stellar hemisphere ($2\pi R^2_\mathrm{*}$) becomes:

\begin{equation}
\frac{\Omega}{2\pi R^2_\mathrm{*}}  = \left(1 - \cos r_\mathrm{spot}\right)\, .
\end{equation}

However, unlike the planet, the starspot emits a flux which means that the flux deficit caused by the starspot, $\Delta F_\mathrm{spot}$ is related to $r_\mathrm{spot}$ by $\rho_\mathrm{spot}$ using the following relation:

\begin{equation}\label{eq:spot_flux}
\Delta F_\mathrm{spot} = \left(1 - \cos r_\mathrm{spot}\right)\left(1 - \rho_\mathrm{spot}\right)
\end{equation}

For the simulations in this work 36 different values of $\rho_\mathrm{spot}$ were used. These were derived using Eq.\,\ref{eq.1} using three values of $T_\mathrm{eff}$, four $T_\mathrm{spot}$ and three $\lambda_\mathrm{obs}$. This equates to 12 different values of $\rho_\mathrm{spot}$ for each stellar spectral class. To determine $\Delta F_\mathrm{spot}$, the mean starspot contrast ($\bar{\rho}_\mathrm{spot}$) was calculated for each spectral class and are given in Table\,\ref{Tab.3.1}.

\begin{table} \centering
\caption{\label{Tab.3.1} The 36 values of $\rho_\mathrm{spot}$ used in the simulations and the three calculated $\bar{\rho}_\mathrm{spot}$ for the M4V, M1V and K5V host star simulations (highlighted in bold).}
\setlength{\tabcolsep}{4pt} \vspace{-5pt}
\begin{tabular}{ccccccccc} 
\hline\hline
 \multicolumn{3}{c}{M4V}  &  \multicolumn{3}{c}{M1V}  &   \multicolumn{3}{c}{K5V}     \\    
\multicolumn{3}{c}{$T_\mathrm{eff} = 3200$\,K}  &  \multicolumn{3}{c}{$T_\mathrm{eff} = 3700$\,K}  &   \multicolumn{3}{c}{$T_\mathrm{eff} = 4100$\,K}     \\  
   $T_\mathrm{spot}$ & $\lambda_\mathrm{obs}$ &  $\rho_\mathrm{spot}$ & $T_\mathrm{spot}$ & $\lambda_\mathrm{obs}$ &  $\rho_\mathrm{spot}$ & $T_\mathrm{spot}$ & $\lambda_\mathrm{obs}$ &  $\rho_\mathrm{spot}$ \\   
 (K) & (nm) &   & (K) & (nm) &   & (K) & (nm) &   \\ 
\hline
           &  600   &  0.61 &          &  600   & 0.56  &          & 600    & 0.53 \\ 
  3000 &  785   &  0.68 & 3400 &  785   & 0.64  & 3700 & 785    & 0.61 \\ 
           &  1000 &  0.74 &          &  1000 & 0.70  &          & 1000  & 0.68 \\ 
\hline  
           &  600   &  0.69 &          &  600   & 0.66  &          & 600    & 0.63 \\ 
  3050 &  785   &  0.75 & 3475 &  785   & 0.72  & 3800 & 785    & 0.70\\ 
           &  1000 &  0.80 &          &  1000 & 0.77  &          & 1000  & 0.75\\ 
\hline  
            &  600  &  0.78 &          &  600   & 0.76  &          & 600    & 0.74 \\ 
  3100 &  785   &  0.83 & 3550 &  785   & 0.81  & 3900 & 785    & 0.79\\ 
           &  1000 &  0.86 &          &  1000 & 0.85  &          & 1000  & 0.83\\ 
\hline  
           &  600   &  0.89 &          &  600   & 0.87  &          & 600    & 0.86 \\ 
  3150 &  785   &  0.91 & 3625 &  785   & 0.90  & 4000 & 785    & 0.89\\ 
           &  1000 &  0.93 &          &  1000 & 0.92  &          & 1000  & 0.91\\ 
\hline   
            &       &  {\bf 0.79} &          &   & {\bf 0.76}  &          &   &{\bf 0.74}\\
\hline \end{tabular} 
\end{table}

\begin{figure*} \includegraphics[width=1.0\textwidth,angle=0]{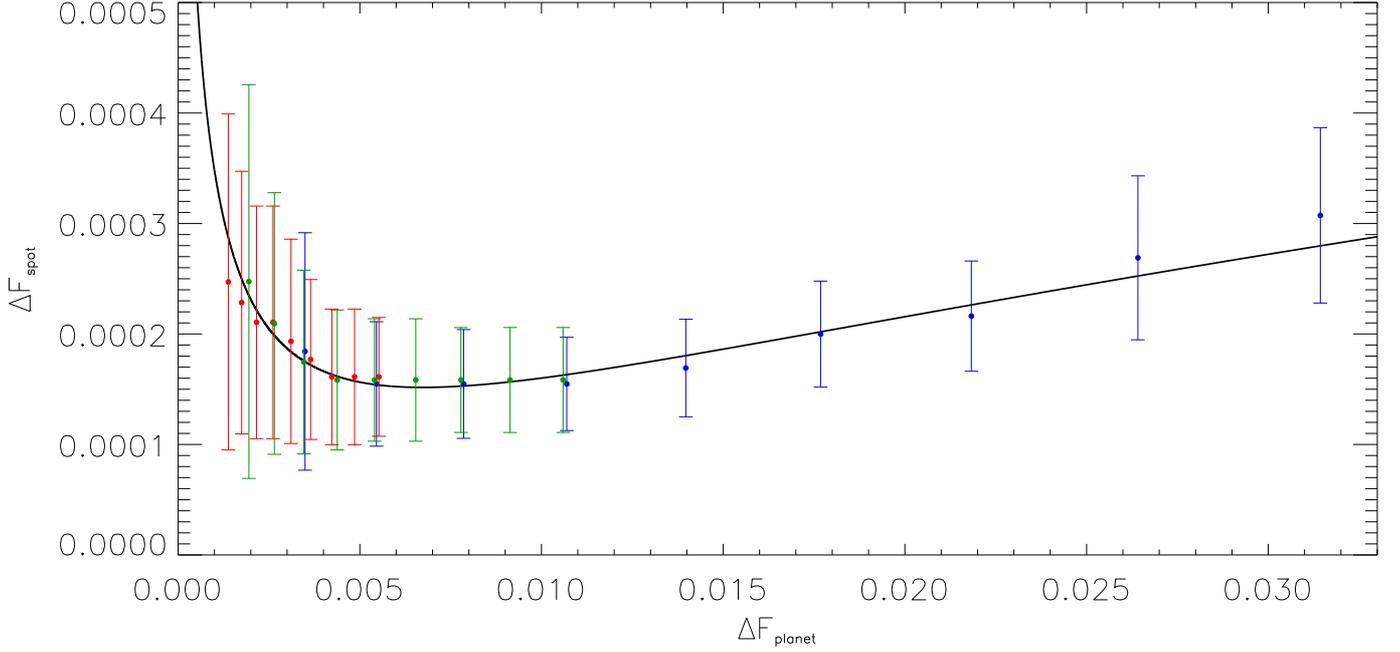} 
\caption{\label{fig:12} Plot of $\Delta F_\mathrm{p}$ and $\Delta F_\mathrm{spot}$ calculated from Eq.\,\ref{Eq.5}\,\&\,\ref{eq:spot_flux} combined with the best fitting model (solid black line). The blue data points represent the M4V host star results. The red data points represent the M1V host star results, while, the green data points represent the K5V host star results.} \end{figure*}

\begin{figure} \includegraphics[width=0.48\textwidth,angle=0]{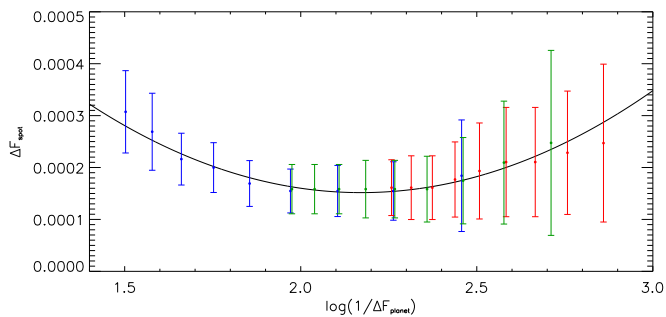} 
\caption{\label{fig:12b} Plot of $\log_{10}\left(\frac{1}{\Delta F_\mathrm{p}}\right)$ and $\Delta F_\mathrm{spot}$ calculated from Eq.\,\ref{Eq.5}\,\&\,\ref{eq:spot_flux} combined with the best fitting model (solid black line). The blue data points represent the M4V host star results. The red data points represent the M1V host star results, while, the green data points represent the K5V host star results.} \end{figure}

Therefore $\Delta F_\mathrm{spot}$ was calculated for the nine $\bar{r}_\mathrm{spot}$ (Table\,\ref{Tab.3}) for each spectral class using the associated $\bar{\rho}_\mathrm{spot}$ (Table\,\ref{Tab.3.1}). The resultant calculated values of $\Delta F_\mathrm{p}$ (Eq.\,\ref{Eq.5}) and $\Delta F_\mathrm{spot}$ (Eq.\,\ref{eq:spot_flux}) are given in Fig.\ref{fig:12}\,\&\,\ref{fig:12b}.


Fig.\,\ref{fig:12} shows that the results for each of the three host stars are in excellent agreement with each other in terms of the smallest $\Delta F_\mathrm{spot}$ that can be detected. The simulation results indicate that when $k\geq 0.10$ then a starspot will need a larger $\Delta F_\mathrm{spot}$ to be detected, either by increasing $r_\mathrm{spot}$ or $\rho_\mathrm{spot}$. The positive gradient towards larger $\Delta F_\mathrm{p}$ represents that larger planets can only detect larger or cooler starspots. The steep negative gradient towards smaller $\Delta F_\mathrm{p}$ shows that the effect from using a fixed cadence combined with the method that is used to detect starspot anomalies, has a stronger influence on the size of starspot which can be detected in TESS transit light curves.


To determine which $\Delta F_\mathrm{p}$ will allow for the smallest $\Delta F_\mathrm{spot}$ to be detected, we fitted a parametrised $2^{nd}$ order polynomial to the results. This was accomplished by parametrising $\Delta F_\mathrm{p}$ to a logarithmic scale:

\begin{equation}\label{eq.5.5}
 t = \log_{10}\left(\frac{1}{\Delta F_\mathrm{p}}\right) \ ,
\end{equation}

\noindent where $t$ is the logarithmic parametrisation of $\Delta F_\mathrm{p}$.

Fig.\,\ref{fig:12b} shows the comparison between $t$ and $\Delta F_\mathrm{spot}$ which indicates a quadratic relation, between the two quantities. We use MCMC to fit a quadratic relation:

\begin{equation}\label{eq.6} 
\begin{split}
\Delta F_\mathrm{spot} = \left(0.00028\pm0.00001\right)t^2 & \\ 
- \left(0.00121\pm0.00001\right)t & \\ 
+ \left(0.00147\pm0.00002\right)
\end{split} 
\end{equation}

Using Eq.\,\ref{eq.6} and setting $\frac{d\,\Delta F_\mathrm{spot}}{d\,t} = 0$ we find that the smallest detectable $\Delta F_\mathrm{spot} = 0.00015\pm0.00001$, when $\Delta F_\mathrm{p} = 0.0067\pm0.0006$ ($k = 0.082\pm0.004$). This allows us to determine that for a 1\Rearth\ planet, a host star with $R_\mathrm{*} = 0.112\pm0.005$\,\Rsun\ will be required to detect starspots with $\Delta F_\mathrm{spot} = 0.00015\pm0.00001$. While for a 3\Rearth\ sub-Neptune planet, a host star with $R_\mathrm{*} = 0.335\pm0.015$\,\Rsun\ will be required to find starspots with $\Delta F_\mathrm{spot} = 0.00015\pm0.00001$.


In Section\,\ref{Sec:2.2} we describe how a degeneracy exists between $r_\mathrm{spot}$ and $\rho_\mathrm{spot}$. For the simulations performed in this work, the smallest $r_\mathrm{spot}$ was determined for sets of fixed parameter values, which included $\rho_\mathrm{spot}$ (calculated from $T_\mathrm{spot}$, $T_\mathrm{eff}$ and $\lambda_\mathrm{obs}$ using Eq.\,\ref{eq.1}). Table\,\ref{Tab.3.1} gives the range of $\rho_\mathrm{spot}$ used in the simulations (0.53--0.93) and when combined with the results given in Tables\,\ref{Tab.M4_Results},\,\ref{Tab.M1_Results}\,and\,\ref{Tab.K5_Results} allows an extrapolation, to determine the largest $\rho_\mathrm{spot}$ for a fixed $r_\mathrm{spot}$.

However, Eq.\,\ref{eq.6} can be used to determine the largest average detectable $\rho_\mathrm{spot}$ for a fixed $r_\mathrm{spot}$. For a 1\,\Rjup\ planet, transiting a 1\,\Rsun\ star, the change in flux caused by the transiting planet is $\Delta F_\mathrm{p} \approx 0.01$. When parametrised using Eq.\,\ref{eq.5.5} gives, $t = 2$. Eq.\,\ref{eq.6} then calculates the smallest detectable $\Delta F_\mathrm{spot}$ for the system, $\Delta F_\mathrm{spot} = 0.000166\pm0.00006$. Using Eq.\,\ref{eq:spot_flux} with a given starspot angular radius, $r_\mathrm{spot} = 5^\circ$ ($r_\mathrm{spot} = 0.087R_\mathrm{*}$) we can determine the lowest contrast detectable in TESS transit light curves for the starspot, $\rho_\mathrm{spot} = 0.96\pm0.02$. While for a $r_\mathrm{spot} = 2^\circ$ ($r_\mathrm{spot} = 0.035R_\mathrm{*}$) we find the lowest contrast detectable to be $\rho_\mathrm{spot} = 0.73\pm0.10$. For a $r_\mathrm{spot} = 1^\circ$ ($r_\mathrm{spot} = 0.017R_\mathrm{*}$) Eq.\,\ref{eq:spot_flux} gives a negative value for $\rho_\mathrm{spot}$, indicating that it is not possible for a $r_\mathrm{spot} = 0.017R_\mathrm{*}$ starspot to be detected in TESS transit data of a 1\,\Rjup\ planet, transiting a 1\,\Rsun\ star.

\section{Discussion and Conclusions} \label{Sec:4}

The key result from the simulations performed in this work indicate that when TESS observes a transiting planet that occults a starspot on the stellar disc, the characteristic ``blip'' of the starspot anomaly will be seen in the resulting light curve. This removes the doubt that starspot anomalies will be present in TESS light curves (due to the TESS primary targets being cooler (redder) host stars with shorter transit durations.) and opens the possibility of using the starspot tracking technique \citep[e.g.,][]{Jeremy2012,Jeremy2015,Mohler2013,Sou2016,Mancini2017} to measure the orbital obliquity of TESS planetary systems. Particularly for targets where the conventional Rossiter-McLaughlin technique \citep{RMa, RMb, RMc, RMd, RMe} is not viable (e.g., slow-rotators, small $k$ and active stars). The simulations then go further in determining the detection limits for a variety of starspot temperatures, planetary \& stellar radii, photometric precision, orbital periods and observational wavelengths. It was found that the starspot detection limits in TESS light curves for M4V, M1V and K5V host stars were $4900\pm1700$\,km, $13800\pm6000$\,km and $15900\pm6800$\,km respectively.

The simulations confirmed expected trends in the detection limits of starspot anomalies, including the photometric parameter dependences of the amplitude of the starspot anomaly. The result confirmed that the amplitude of a starspot anomaly is strongly dependent on $r_\mathrm{spot}$ and $\rho_\mathrm{spot}$, and therefore, by association with $\rho_\mathrm{spot}$, strongly dependent on $T_\mathrm{eff}$ and $\lambda_\mathrm{obs}$. While, the amplitude of the starspot anomaly is less dependent on $P$. The simulations then show that overall the detection limit of a starspot anomaly is dependent on $R_\mathrm{p}$ (Table\,\ref{Tab.3}) and is therefore dependent on, $\Delta F_\mathrm{planet}$. However, the simulations uncovered an unexpected trend: starspots which are small and hot can be detected by large planets, however, they can't be detected by smaller planets.

From our investigations of the synthetic light curves, we uncovered that the unexpected trend was a detection bias, generated by the fixed 2\,min cadence of TESS combined with the method used to determine if the starspot anomaly was detectable, by comparing the mean amplitude of a starspot anomaly to the observational noise. This detection method is in agreement with fitting starspot anomalies in transit light curves. If the mean amplitude of the starspot anomaly is not large enough, it is not possible to constrain the starspot properties (see Section\,\ref{Sec:3}) It was found that for anomalies with a short maximum duration (apex shape) the data points lying on the ingress and egress of the starspot anomaly reduced the mean amplitude. As a consequence for the starspot to be detected either $r_\mathrm{spot}$ or $R_\mathrm{p}$ needed to be increased to increase the mean amplitude. Though as in the case depicted in Fig.\,\ref{fig:4}, if the maximum amplitude of the starspot anomaly is below the detection threshold, then increasing $r_\mathrm{spot}$ further did not allow a detection. If instead of using the mean amplitude, the maximum amplitude was tested against the rms scatter in the data, it would allow starspot anomalies described by a single data point (i.e., apex shape) to be considered detected, when in reality would be dismissed as noise (e.g., Fig.\,\ref{fig:9}). By using the mean amplitude of the starspot anomaly, effectively mimics what happens when a transit-starspot model fits a starspot anomaly. Hence, the unexpected trend -- smaller hot starspots will be detected by larger planets -- will be present when analysing TESS transit light curves. To counter this would require a reduction in the observational cadence, which would increase the number of data points which describe the starspot anomaly. However, this is not practical when analysing TESS transit data and would require further follow-up photometric observations using a different telescope.

By separating the simulations into three groups for each of the three host stars and then converting $R_\mathrm{p}$ and $\bar{r}_\mathrm{spot}$ into $\Delta F_\mathrm{planet}$ and $\Delta F_\mathrm{spot}$, a direct comparison between the planets and starspots for the three host stars was made. The resultant comparison between $\Delta F_\mathrm{planet}$ and $\Delta F_\mathrm{spot}$ showed excellent agreement of the simulations between the three host stars and indicate that the smallest and hottest starspots detected in TESS transit light curves will be when $k = 0.082\pm0.004$. By fitting a parametrised $2^{nd}$ order polynomial to the results, we are able to determine the minimum size and lowest contrast (highest intensity) that can be detected for a given value of $k$ in the TESS exoplanetary transit data. The comparison also revealed that the effect from using a fixed 2\,min cadence combined to using the mean flux of a starspot anomaly to determine its detection had a larger impact on $\Delta F_\mathrm{spot}$ than $R_\mathrm{p}$.

Young K and M dwarf stars are known to have a high starspot surface coverage \citep[e.g.,][]{Oneal2004, Jackson2013}. However, the activity-age relationship is still unknown. This leads to the question, how many TESS transit light curves will contain starspot anomalies. The simulations presented in this work are the starting point, in beginning to answer this question. In that the simulations show that starspot anomalies will be observable in TESS transit light curves. The starspot distribution on the stellar disc will also be an important factor in this question. \citet{Barnes_J2015} used high resolution Doppler images of two M dwarfs (GJ\,791.2\,A; M4.5: LP\,944-20; M9) to determine the overall starspot filling factors $\approx3$\,\% and $\approx1.5$\,\% for the two M dwarfs. However, \citet{Barnes_J2015} found maximum starspot filling factors $82.3$\,\% and $76.6$\,\% at high latitudes (polar). If the starspot distributions of GJ\,791.2\,A and LP\,944-20 can be considered typical of M dwarf stars that will be observed by TESS. Then a polar biased starspot distribution will affect the total number of TESS transit light curves, which contain a starspot anomaly. By determining the smallest detectable starspot size for a range of planetary sizes, starspot temperatures, orbital periods and noise levels is an essential first step in beginning to forecast the number of TESS transit light curves which will be affected by starspot anomalies. Therefore, gauging the importance and the community future reliance on transit-starspot models.


\begin{acknowledgements}
      We would like to thank the anonymous referee and the editor for their helpful comments, which improved the quality of this manuscript. This work was supported by a CONICYT / FONDECYT 2018 Postdoctoral research grant, project number: 3180071. JTR thanks the Centro de Astronom\'{i}a (CITEVA), Universidad de Antofagasta for hosting the CONICYT / FONDECYT 2018 Postdoctoral research grant. The following internet-based resources were used in the research for this paper: the NASA Astrophysics Data System; the SIMBAD database and VizieR catalogue access tool operated at CDS, Strasbourg, France; and the ar$\chi$iv scientific paper preprint service operated by Cornell University.
\end{acknowledgements}

\begin{appendix}
 
\section{Full results of all simulations}

\longtab[1]{
\begin{longtable}{cccccccccccc} 
\caption{\label{Tab.M4_Results}The smallest detectable value of $r_\mathrm{spot}$ determined from the 6718 simulations of the 1296 scenarios that used an M4V dwarf host star.}\\
\hline\hline
\noalign{\smallskip}
  & & \multicolumn{3}{c}{$\lambda_{obs}=600$\,nm}  &  \multicolumn{3}{c}{$\lambda_{obs}=785$\,nm}  &   \multicolumn{3}{c}{$\lambda_{obs}=1000$\,nm}     \\    
       $T_\mathrm{spot}$ & rms scatter& \multicolumn{3}{c}{$P$}  &  \multicolumn{3}{c}{$P$}  &   \multicolumn{3}{c}{$P$} \\
 ($K$) & (ppm)  & 1\,d & 2\,d &  3\,d & 1\,d & 2\,d &  3\,d & 1\,d & 2\,d &  3\,d \\   
\hline
\noalign{\smallskip}
\multicolumn{11}{c}{$R_\mathrm{p} = 1.00$\Rearth} \\
\hline
3000&      60&0.017$R_\mathrm{*}$&0.026$R_\mathrm{*}$&0.026$R_\mathrm{*}$&0.017$R_\mathrm{*}$&0.026$R_\mathrm{*}$&0.026$R_\mathrm{*}$&0.026$R_\mathrm{*}$&0.026$R_\mathrm{*}$&0.026$R_\mathrm{*}$\\
    3000&     100&0.026$R_\mathrm{*}$&0.026$R_\mathrm{*}$&0.035$R_\mathrm{*}$&0.026$R_\mathrm{*}$&0.035$R_\mathrm{*}$&0.035$R_\mathrm{*}$&0.026$R_\mathrm{*}$&0.035$R_\mathrm{*}$&0.044$R_\mathrm{*}$\\
    3000&     150&0.026$R_\mathrm{*}$&0.035$R_\mathrm{*}$&0.044$R_\mathrm{*}$&0.035$R_\mathrm{*}$&0.035$R_\mathrm{*}$&0.044$R_\mathrm{*}$&0.035$R_\mathrm{*}$&0.044$R_\mathrm{*}$&0.044$R_\mathrm{*}$\\
\hline
\end{longtable}
}

\longtab[2]{\begin{scriptsize}\begin{landscape}
\begin{longtable}{cccccccccccc} 
\caption{\label{Tab.M1_Results}The smallest detectable value of $r_\mathrm{spot}$ determined from the 7812 simulations of the 1296 scenarios that used an M1V dwarf host star. For a direct comparison with the results from the M4V dwarf host star simulations, the values of $r_\mathrm{spot}$ in brackets are the starspot size if they were on the M4V dwarf.}\\
\hline\hline
\noalign{\smallskip}
  & & \multicolumn{3}{c}{\normalsize $\lambda_{obs}=600$\,nm}  &  \multicolumn{3}{c}{\normalsize$\lambda_{obs}=785$\,nm}  &   \multicolumn{3}{c}{\normalsize$\lambda_{obs}=1000$\,nm}     \\    
      {\normalsize $T_\mathrm{spot}$} & {\normalsize rms scatter}& \multicolumn{3}{c}{\normalsize$P$}  &  \multicolumn{3}{c}{\normalsize$P$}  &   \multicolumn{3}{c}{\normalsize$P$} \\
 {\normalsize($K$)} & {\normalsize(ppm)}  & {\normalsize2\,d} & {\normalsize3\,d} &  {\normalsize4\,d} & {\normalsize2\,d} & {\normalsize3\,d} &  {\normalsize4\,d} & {\normalsize2\,d} & {\normalsize3\,d} &  {\normalsize4\,d} \\   
\hline
\noalign{\smallskip}
\multicolumn{11}{c}{\normalsize$R_\mathrm{p} = 2.00$\Rearth} \\
\hline

3400 &60 &0.017$R_\mathrm{*}$\,\textcolor{red}{(0.056$R_\mathrm{*}$)} &0.026$R_\mathrm{*}$\,\textcolor{red}{(0.083$R_\mathrm{*}$)} &0.026$R_\mathrm{*}$\,\textcolor{red}{(0.083$R_\mathrm{*}$)} &0.017$R_\mathrm{*}$\,\textcolor{red}{(0.056$R_\mathrm{*}$)} &0.026$R_\mathrm{*}$\,\textcolor{red}{(0.083$R_\mathrm{*}$)} &0.026$R_\mathrm{*}$\,\textcolor{red}{(0.083$R_\mathrm{*}$)} &0.026$R_\mathrm{*}$\,\textcolor{red}{(0.083$R_\mathrm{*}$)} &0.026$R_\mathrm{*}$\,\textcolor{red}{(0.083$R_\mathrm{*}$)} &0.026$R_\mathrm{*}$\,\textcolor{red}{(0.083$R_\mathrm{*}$)} \\
\hline
\end{longtable}\end{landscape}\end{scriptsize}
}

\longtab[3]{\begin{scriptsize}\begin{landscape}
\begin{longtable}{cccccccccccc} 
\caption{\label{Tab.K5_Results}The smallest detectable value of $r_\mathrm{spot}$ determined from the 6737 simulations of the 1296 scenarios that used an K5V dwarf host star. For a direct comparison with the results from the M4V dwarf host star simulations, the values of $r_\mathrm{spot}$ in brackets are the starspot size if they were on the M4V dwarf.}\\
\hline\hline
\noalign{\smallskip}
  & & \multicolumn{3}{c}{\normalsize $\lambda_{obs}=600$\,nm}  &  \multicolumn{3}{c}{\normalsize$\lambda_{obs}=785$\,nm}  &   \multicolumn{3}{c}{\normalsize$\lambda_{obs}=1000$\,nm}     \\    
      {\normalsize $T_\mathrm{spot}$} & {\normalsize rms scatter}& \multicolumn{3}{c}{\normalsize$P$}  &  \multicolumn{3}{c}{\normalsize$P$}  &   \multicolumn{3}{c}{\normalsize$P$} \\
 {\normalsize($K$)} & {\normalsize(ppm)}  & {\normalsize2\,d} & {\normalsize4\,d} &  {\normalsize6\,d} & {\normalsize2\,d} & {\normalsize4\,d} &  {\normalsize6\,d} & {\normalsize2\,d} & {\normalsize4\,d} &  {\normalsize6\,d} \\   
\hline
\noalign{\smallskip}
\multicolumn{11}{c}{\normalsize$R_\mathrm{p} = 3.00$\Rearth} \\
\hline
3700 &60 &0.017$R_\mathrm{*}$\,\textcolor{red}{(0.070$R_\mathrm{*}$)} &0.017$R_\mathrm{*}$\,\textcolor{red}{(0.070$R_\mathrm{*}$)} &0.026$R_\mathrm{*}$\,\textcolor{red}{(0.105$R_\mathrm{*}$)} &0.017$R_\mathrm{*}$\,\textcolor{red}{(0.070$R_\mathrm{*}$)} &0.017$R_\mathrm{*}$\,\textcolor{red}{(0.070$R_\mathrm{*}$)} &0.026$R_\mathrm{*}$\,\textcolor{red}{(0.105$R_\mathrm{*}$)} &0.017$R_\mathrm{*}$\,\textcolor{red}{(0.070$R_\mathrm{*}$)} &0.026$R_\mathrm{*}$\,\textcolor{red}{(0.105$R_\mathrm{*}$)} &0.026$R_\mathrm{*}$\,\textcolor{red}{(0.105$R_\mathrm{*}$)} \\
\hline

\end{longtable}\end{landscape}\end{scriptsize}
}

\end{appendix}

\end{document}